\documentclass[11pt]{salam}

\usepackage{physics}
\usepackage{graphicx}
\usepackage{cancel}
\usepackage{amsmath}
\usepackage{amssymb}

\renewcommand{\var}{\operatorname{var}}

\def\T{{\operatorname{T}}}

\def\cov{{\operatorname{cov}}}
\def\corr{{\operatorname{corr}}}
\graphicspath{{img/}}

\title[Dynamical Mean-Field Theory Tutorial]{
  Building Intuition for Dynamical Mean-Field Theory: A Simple Model and the Cavity Method
}
\subtitle{A tutorial oriented to the biophysics community}
\author{Emmy Blumenthal}
\affiliation{Princeton University Department of Physics, Princeton, NJ 08540, USA}
\date{\today}

\begin{document}
\maketitle
% \margintoc
\begin{abstract}
  Dynamical Mean-Field Theory (DMFT) is a powerful theoretical framework for analyzing systems with many interacting degrees of freedom. This tutorial provides an accessible introduction to DMFT. We begin with a linear model where the DMFT equations can be derived exactly, allowing readers to develop clear intuition for the underlying principles. We then introduce the cavity method, a versatile approach for deriving DMFT equations for non-linear systems. The tutorial concludes with an application to the generalized Lotka--Volterra model of interacting species, demonstrating how DMFT reduces the complex dynamics of many-species communities to a tractable single-species stochastic process. Key insights include understanding how quenched disorder enables the reduction from many-body to effective single-particle dynamics, recognizing the role of self-averaging in simplifying complex systems, and seeing how collective interactions give rise to non-Markovian feedback effects.
\end{abstract}
\tableofcontents

\paragraph{A brief note of acknowledgment.} Much of my understanding of this content comes from research studying ecological models. I learned many of these ideas from references~\cite{Nishimori,cui2024leshoucheslecturescommunity, Roy_2019, cuiPRM, bunin, cui2020effect,potters2020first, Zou_Huang_2024}. Of course, all errors are my own.
I am grateful to Akshit Goyal, Gautam Reddy, and Pankaj Mehta for their feedback on these notes.
Please contact me if you identify any errors or have suggestions for improvement.\sidenote{eblu $[\alpha \tau]$ princeton.edu}

\newpage
\section{Motivation}

Dynamical mean-field theory (DMFT) is a powerful theoretical framework to reduce the equations of motion for a dynamical system with many degrees of freedom to a single equation of motion for a single degree of freedom, sometimes called the `mean-field' particle.
DMFT was originally developed to study the dynamics of spin glasses but has since been applied to a wide variety of systems that involve dense and disordered interactions between many heterogeneous degrees of freedom.
For example, in ecology, we might be interested in dynamics of the abundances of species in a community where each species interacts with all other species.
A model for this system might look like:
\begin{equation}
  \begin{aligned}
    \dv{N_i}{t}
    &=
    N_i \qty(
      1
      -
      N_i
      -
      \sum_{j(\neq i)}
      A_{ij} N_j
    )
    ,
    \qquad
    i = 1, \ldots, S
  \end{aligned}
\end{equation}
where $N_i$ is the abundance of species $i$ and $A_{ij}$ is a matrix of interaction parameters between species $i$ and $j$.
This model is called the generalized Lotka--Volterra (GLV) model of population dynamics.
Making definitive statements about the dynamics of this entire system in the $S \gg 1$ limit is difficult because it is non-linear and involves $O(S^2)$ interaction parameters.
To avoid this problem, in the $S \gg 1$ limit (the thermodynamic limit), we might consider parameters $A_{ij}$ to be drawn from some distribution $P(\{A_{ij}\})$ and notice that species $i$ is interacting with a sum over many other species (a `mean field').
We might replace such a sum with a stochastic process and get an equation of motion for just species $i$ of the form:
\begin{equation}
  \begin{aligned}
    \dv{N_i}{t}
    &=
    \text{some expression only in terms of $N_i$}
    .
    \label{eq:GLV-mystery-equation}
  \end{aligned}
\end{equation}
Such an equation of motion is called the DMFT equation of motion.
Deriving this sort of equation is advantageous because it allows us to apply tools from the theory of stochastic process to study the dynamics of the (deterministic) system, potentially facilitating more tractable theoretical analysis.
At the end of this note in section~\ref{sec:generalized-lotka-volterra}, we will derive the right-hand side of Eq.~\eqref{eq:GLV-mystery-equation}.

Another example application of DMFT is the study of neural networks which are some of the first biophysical subjects of study to which DMFT was applied \cite{SompolinskyPathIntegral,SompolinskyPRL}.
A model for neural dynamics might describe the activations $\{S_i\}$ of $N$ neurons which each experience a local field $h_i$ so that $S_i = \phi(h_i)$ with $\phi$ being a non-linear gain function (for example $\phi(x) = \tanh(g x)$).
The local fields undergo dynamics:
\begin{equation}
  \begin{aligned}
    \dv{h_i}{t}
    &=
    -h_i
    +
    \sum_{j=1}^N
    J_{ij} S_j(t),
    \qquad
    i = 1, \ldots, N,
  \end{aligned}
\end{equation}
which models the local field $h_i$ as a sum of contributions from all other neurons $S_j$ weighted by some synaptic weights $J_{ij}$.
A DMFT approach to this system would be to treat $J_{ij}$ as a random matrix drawn from some distribution $P(\{J_{ij}\})$ and then derive an equation of motion for a single, randomly chosen neuron.
This allows for analysis of many behaviors without having to specify $O(N^2)$ synaptic weights.

In appendix~\ref{sec:brief-intro-rmt}, we will briefly review random matrix theory (RMT) which will serve as a useful tool for deriving the DMFT equations.
In appendix~\ref{sec:brief-review-stationary-gaussian-processes}, we will briefly review stationary Gaussian processes which will appear in the DMFT equations.

\section{Exactly deriving the DMFT of a linear model\label{sec:exactly-deriving-dmft-linear-model}}

In this section, we will exactly derive the DMFT equations for a linear model.
The techniques we will apply here are typically not possible in any non-linear model because we can exactly solve the linear model.
However, this example is useful because it allows us to precisely identify what assumptions we make when deriving a DMFT.
It also clarifies why certain steps we take in a method like the cavity method are justified.
It's important that a reader who has not been exposed to DMFT before knows that \textbf{the approach we take in this section is almost never possible for any setup except linear models}.
An approach that works more generally is the cavity method, which we will discuss in section~\ref{sec:cavity-method}.
Nonetheless, I believe this example gets to the heart of what DMFT is and why it is useful.

\begin{figure}[p]
  \begin{widetext}
  \centering
  \includegraphics[]{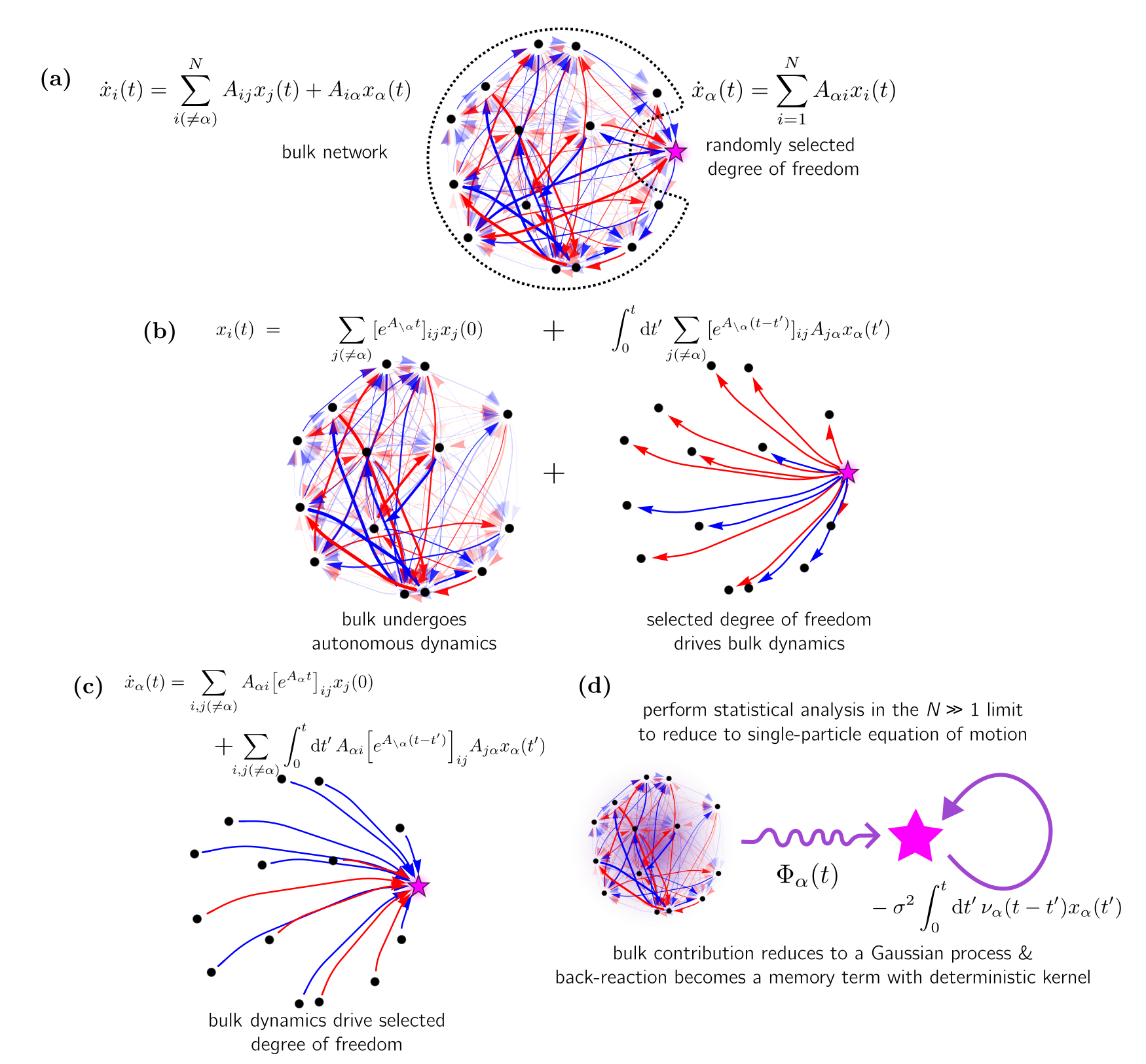}
  \caption{\label{fig:solvable-dmft-schematic}
      Schematic of DMFT derivation presented in section~\ref{sec:exactly-deriving-dmft-linear-model}.
      (a)
      Fully connected linear network with antisymmetric interactions ($A_{ij} = -A_{ji}$). Blue and red arrows represent positive and negative interactions, respectively.
      A randomly selected degree of freedom (star) is chosen and treated separately from the rest of the network.
      (b)
      Solve the dynamics for $x_i$ without $x_\alpha$.
      There are two linear contributions: the autonomous (self-driven) dynamics of the network without $x_\alpha$ and the dynamics driven by the dynamics of $x_\alpha$.
      (c)
      The dynamics of $x_\alpha$ are driven by the dynamics of all other $x_i$
      (d)
      After statistically analyzing the resulting equation of motion for $x_\alpha$, one finds two linear contributions. One is due to the autonomous dynamics of all other $x_i$ and converges to a Gaussian process in the large-$N$ limit. The other is due to a back-reaction from $x_\alpha$ driving the rest of the system and the bulk driving $x_\alpha$ back; it becomes a memory term in the large-$N$ limit with a deterministic kernel.
  }
  \end{widetext}
\end{figure}

Figure~\ref{fig:solvable-dmft-schematic} shows a schematic of the steps we will take to derive the DMFT equations for this linear model.

\subsection{The model}

We will consider a linear model of the form:
\begin{equation}
  \begin{aligned}
    \dv{x_i}{t}
    &=
    \sum_{j=1}^N
    A_{ij}
    x_j(t),
    \qquad
    i = 1, \ldots, N,
  \end{aligned}
\end{equation}
where $A_{ij}$ is an $N \times N$ skew-symmetric $(A_{ij} = -A_{ji})$ matrix.
We will assume that the entries of the matrix $A_{ij}$ are drawn from a Gaussian distribution with statistics:\sidenote{It is not strictly necessary to assume that $A_{ij}$ follow a normal distribution, but it simplifies the analysis.}
\begin{equation}
  \begin{aligned}
    \ev{A_{ij}} &= 0,
    \qquad
    \ev{A_{ij} A_{kl}} = 
    \frac{\sigma^2 }{N}
    \qty(
      \delta_{ik} \delta_{jl}
      -
      \delta_{il} \delta_{jk}
    )
  \end{aligned}
\end{equation}
where $\sigma^2$ is a constant that sets the scale of the interactions and $\delta_{ij}$ is the Kronecker delta symbol (1 if $i = j$, 0 otherwise).
This says that $A_{ij}$ are independent up to the constraint that $A_{ij} = -A_{ji}$, have zero mean, and have variance $\sigma^2/N$.
We scale the variance of $A_{ij}$ by $1/N$ so that the time derivative of $x_i$ is of order 1 in the thermodynamic limit $N \gg 1$.
We will similarly assert that the initial conditions $x_1,\ldots,x_N$ are drawn independently from a Gaussian distribution with zero mean and variance $\sigma_0^2$:
\begin{equation}
  \begin{aligned}
    \ev{x_i(0)} = 0,
    \qquad
    \ev{x_i(0) x_j(0)} =
    \sigma_0^2 \,\delta_{ij}.
  \end{aligned}
\end{equation}
% \begin{equation}
%   \begin{aligned}
%     \var\qty[\dv{x_i}{t}]
%     &=
%     \sum_{j,k=1}^N
%     \ev{A_{ij} A_{ik}}
%     x_j x_k
%     % \\
%     % &
%     =
%     \frac{\sigma^2}{N}
%     \sum_{j=1}^N
%     x_j^2
%     =
%     O(1).
%   \end{aligned}
% \end{equation}

\subsection{Choosing a random particle: two types of randomness}

The goal of DMFT is to reduce the equations of motion for the $N$ particles to a single equation of motion for a single particle.
To do this, we choose a random particle which we will label $i = \alpha$ and then eliminate all other particles.
In our simple example, we will eliminate all other particles by solving their equations of motion in terms of the random particle $\alpha$ and then substituting this solution into the equation of motion for $\alpha$.
Before we do this, we need to explicitly identify that there are two types of randomness in our model.
There is the randomness due to sampling the entries of the matrix $A_{ij}$ and initial conditions $x_i(0)$ which is called \textit{quenched disorder}.
There is also the randomness due to the choice of the random particle $\alpha$, which I will call \textit{shuffling randomness}.
To be explicit, an average of a quantity $X$ over quenched disorder is defined:
\begin{equation}
  \begin{aligned}
    \ev{X}_\mathcal{Q}
    =
    \int \prod_{i,j=1}^N \dd{A_{ij}}
    \prod_{i=1}^N \dd{x_i(0)}
    P(\{A_{ij}\}, \{x_i(0)\})
    \, X,
  \end{aligned}
\end{equation}
and an average over shuffling randomness is defined:
\begin{equation}
  \begin{aligned}
    \ev{X}_\text{sh}
    =
    \frac{1}{N}
    \sum_{\alpha=1}^N
    \qty(
      \text{$X$ with $\alpha$ being the chosen random particle}
    ).
  \end{aligned}
  \label{eq:shuffling-average-definition}
\end{equation}
The goal of DMFT is to derive a stochastic differential equation which has some noise term in which sampling the noise is equivalent to shuffling randomness.\sidenote{
The term `shuffling randomness' is not standard, but I find it useful to distinguish the two types of randomness.
It's more standard to call the sort of average in Eq.~\eqref{eq:shuffling-average-definition} a single-realization system average.
}
That is to say: the dynamics of every particle are equivalent to the dynamics of a single particle with some noise term which is `resampled' for each particle.
Such an interpretation is useful because it allows us to use tools from the theory of stochastic processes even for a single, frozen realization of the quenched disorder (the matrix $A_{ij}$ and initial conditions $x_i(0)$).

\subsection{Integrating out $N - 1$ degrees of freedom}

We will now proceed to derive the DMFT equations for the random particle $\alpha$.
We do this by solving the equations of motion for the other $N - 1$ particles in terms of the random particle $\alpha$ and then substituting this solution into the equation of motion for $\alpha$.
We split the equation of motion for the random particle $\alpha$ into two parts:
\begin{equation}
  \begin{aligned}
    \dv{x_i}{t}
    &=
    \sum_{j(\neq \alpha)}
    A_{ij}
    x_j(t)
    +
    A_{i\alpha}
    x_\alpha(t),
    \quad
    (i \neq \alpha),
    \\
    \dv{x_\alpha}{t}
    &=
    \sum_{i (\neq \alpha)}
    A_{\alpha i}
    x_i(t)
    +
    A_{\alpha \alpha}
    x_\alpha(t).
  \end{aligned}
\end{equation}
We can exactly solve the equations of motion for the $N - 1$ particles $i \neq \alpha$ in terms of the random particle $\alpha$:\sidenote{
This is accomplished by first writing the equations of motion for the $N - 1$ particles in matrix form:
\begin{align*}
  \dv{\mathbf{x}_{\setminus \alpha}}{t}
  &=
  A_{\setminus \alpha} \mathbf{x}_{\setminus \alpha}(t)
  +
  A_{\setminus \alpha, \alpha} x_\alpha(t),
\end{align*}
where $\mathbf{x}_{\setminus \alpha}$ is the vector with the $\alpha$th component removed.
We then note that $\mathbf{x}_{\setminus \alpha}(t) = e^{A_{\setminus \alpha} t} \mathbf{x}_{\setminus \alpha}(0)$ solves the equation of motion if we neglect the term involving $x_\alpha(t)$.
Finally, we use that the equation is linear to get Eq.~\eqref{eq:xi_solution}.
}
\begin{equation}
  \label{eq:xi_solution}
  \begin{aligned}
    x_i(t)
    &=
    \sum_{j(\neq \alpha)}
    [
      e^{A_{\setminus \alpha} t}
    ]_{ij}
    x_j(0)
    +
    \int_0^t
    \dd{t'}
    \sum_{j(\neq \alpha)}
    [
      e^{A_{\setminus \alpha} (t - t')}
    ]_{ij}
    A_{j\alpha}
    x_\alpha(t')
  \end{aligned}
\end{equation}
where we have introduced the notation $A_{\setminus \alpha}$ to denote the matrix $A$ with the row and column corresponding to the random particle $\alpha$ removed.
Here, $e^{A_{\setminus \alpha} t}$ is the matrix exponential.
We then substitute this solution into the equation of motion for the random particle $\alpha$:
\begin{equation}
	\begin{aligned}
		\dv{x_\alpha}{t}
		&=
		\Phi_\alpha(t)
		-
		\sigma^2
		\int_0^t
		\dd{t'}
		\nu_\alpha(t-t')
		x_\alpha(t'),
	\end{aligned}
  \label{eq:dmft-eom-before-stat}
\end{equation}
where we have defined:
\begin{equation}
	\begin{aligned}
		\Phi_\alpha(t)
		&\equiv
		\sum_{i,j(\neq \alpha)}
		A_{\alpha i}
		[
			e^{A_{\setminus \alpha} t}
		]_{ij}
		x_j(0),
		\\
		\nu_\alpha(t)
		&\equiv
		-
		\frac{1}{\sigma^2}
		\sum_{i,j(\neq \alpha)}
		A_{\alpha i}
		[
			e^{A_{\setminus \alpha} t}
		]_{ij}
		A_{j\alpha}
    .
	\end{aligned}
  \label{eq:Phi-nu-definitions}
\end{equation}
Note that we have dropped $A_{\alpha\alpha}$ because it is zero from the assumption that $A_{ij}$ is skew-symmetric.\sidenote{Generally, you can drop these sorts of terms in DMFT because they are of order $1/\sqrt{N}$ and thus negligible in the thermodynamic limit.}
This constitutes the first step in deriving the DMFT equation of motion.
This result, however, is not yet in a form that is useful for analysis because we don't know much about $\Phi_\alpha(t)$ and $\nu_\alpha(t)$.

\subsection{Statistical analysis and self-averaging}

\subsubsection{$\Phi_\alpha$ fluctuations under quenched disorder at fixed $\alpha$}

To get a more useful result, we look to the statistics of the random variables $\Phi_\alpha(t)$ and $\nu_\alpha(t)$.
Recall that $\Phi_\alpha(t)$ and $\nu_\alpha(t)$ may fluctuate over both quenched disorder and shuffling randomness.
We will first treat the quenched disorder and then see how it interacts with the shuffling randomness.
This means that we will first consider the case where $\alpha$ is fixed.
Under this assumption, we might notice that $\Phi_\alpha(t)$ is a sum of $\{A_{\alpha i}\}$ which are independent Gaussian random variables which means that under fluctuations over quenched disorder, $\Phi_\alpha(t)$ is also a Gaussian process.\sidenote{If you are uncomfortable with this argument, I've included a brief review of stationary Gaussian processes in appendix~\ref{sec:brief-review-stationary-gaussian-processes}.}
A Gaussian process is completely characterized by its first two moments, the mean and covariance:
% \begin{widetext}
\begin{gather}
  \begin{aligned}
    \ev{\Phi_\alpha(t)}_\mathcal{Q}
  &=
  \sum_{i,j(\neq \alpha)}
  \ev{A_{\alpha i}}
  \ev{[e^{A_{\setminus \alpha} t}]_{ij}}
  \cancel{\ev{x_j(0)}_\mathcal{Q}}
  =
  0,
  \end{aligned}
  \label{eq:Phi-quenched-mean}
  \\
  \begin{aligned}
    \ev{\Phi_\alpha(t)\Phi_\alpha(s)}_\mathcal{Q}
  &=
  \sum_{i,j,k,l(\neq \alpha)}
  \ev{A_{\alpha i} A_{\alpha k}}
  \ev{[e^{A_{\setminus \alpha} t}]_{ij} [e^{A_{\setminus \alpha} s}]_{kl}}
  \ev{x_j(0)x_l(0)}
  \\
  &
  =
  \sigma^2
  \sigma_0^2
  \frac{1}{N}
  \sum_{i,j(\neq \alpha)}
  \ev{[e^{A_{\setminus \alpha} t}]_{ij} [e^{A_{\setminus \alpha} s}]_{ij}}
  \\
  &
  \approx
  \sigma^2
  \sigma_0^2
  \ev{
  \frac{1}{N}
  \Tr
  \qty[e^{A_{\setminus \alpha} t + A_{\setminus \alpha}^\T s}]
  }
  \\
  &=
  \sigma^2
  \sigma_0^2
  \ev{
  \frac{1}{N}
  \sum_{i(\neq \alpha)}
  e^{
    \lambda_i t + \lambda_i^\ast s
  }
  }
  .
  \end{aligned}
  \label{eq:Phi-quenched-covariance}
\end{gather}
% \end{widetext}
Notice that in Eq.~\eqref{eq:Phi-quenched-covariance}, we were able to factor the averages because the entries of the matrix $A_{\setminus \alpha}$ are (definitionally) independent of $A_{\alpha i}$.
Here, $\lambda_i$ are the eigenvalues of the matrix $A_{\setminus \alpha}$.
In the $N \gg 1$ limit, we can replace the average over the eigenvalues with an integral over the spectral density of the matrix $A$:\sidenote{We can use the spectral density of $A$ instead of $A_{\setminus \alpha}$ because they are the same up to a small correction of order $1/N$ in the $N \gg 1$ limit.}
\begin{equation}
  \begin{aligned}
    \frac{1}{N}
    \sum_{i(\neq \alpha)}
    e^{
      \lambda_i t + \lambda_i^\ast s
    }
    \to
    \int \dd[2]{\lambda}
    \rho_{A}(\lambda)
    \,
    e^{
      \lambda t + \lambda^\ast s
    }
  \end{aligned}
\end{equation}
where:
\begin{equation}
  \begin{aligned}
    \rho_A(\lambda)
    =
    \frac{1}{N}
    \sum_{i=1}^N
    \delta(x - \lambda_i)
  \end{aligned}
\end{equation}
is the empirical distribution of the eigenvalues of the matrix $A$, called the spectral density.
In the thermodynamic limit this becomes:
\begin{equation}
	\begin{aligned}
		\rho_A(x+i y)
		&=
		\begin{cases}
			\delta(x)
			\cdot
			\frac{1}{2\pi\sigma^2}
			\sqrt{
				4 \sigma^2 - y^2
			},
			& \text{if } -2\sigma \leq y \leq 2\sigma, \\
			0,
			& \text{otherwise,}
		\end{cases}
	\end{aligned}
\end{equation}
which is the Wigner semicircle law but rotated onto the imaginary axis because $A_{ij}$ is skew-symmetric and thus has purely imaginary eigenvalues.\sidenote{
If you are unfamiliar with these concepts, I have included a brief review of random matrix theory in appendix~\ref{sec:brief-intro-rmt}.
}
We can exploit these results to write:
\begin{equation}
  \begin{aligned}
    \ev{\Phi_\alpha(t)\Phi_\alpha(s)}_\mathcal{Q}
    &=
    \frac{\sigma_0^2}{2\pi}
		\int_{-2\sigma}^{2\sigma}
		\dd{y}
		\sqrt{4\sigma^2 - y^2}
    \,
		e^{i y(t-s)}
    \\
    &
		=
		\sigma_0^2
		\sigma^2
		\,
		\frac{J_1(2 \sigma |t-s|)}{\sigma|t-s|},
  \end{aligned}
\end{equation}
where $J_1$ is the Bessel function of the first kind.\sidenote{
The Bessel function of the first kind is defined as:
\begin{align*}
J_1(x)
=
\frac{1}{\pi}
\int_0^\pi
\cos(\theta - x \sin \theta)
\dd{\theta}
.
\end{align*}
It has the $x \to \infty$ asymptotic behavior:
\begin{align*}
  J_1(x)
  &\sim
  \sqrt{
    \frac{2}{\pi x}
  }
  \sin(x - \tfrac{\pi}{4})
  .
\end{align*}
}

\subsubsection{Fluctuations of $\Phi_\alpha$ under shuffling randomness}

Now, we might want to consider how $\Phi_\alpha(t)$ fluctuates under shuffling randomness.
To do this, ,,we might try to compute $\ev{\Phi_{\alpha}(t)}_\text{sh}
=
N^{-1}\sum_{\alpha=1}^N \Phi_\alpha(t)
$ and $\ev{\Phi_\alpha(t)\Phi_\alpha(s)}_\text{sh} = N^{-1}\sum_{\alpha=1}^N \Phi_\alpha(t)\Phi_\alpha(s)$.
However, there's not really much progress we can make because whatever expression we get will still involve a sum over $\alpha$ which is very difficult to compute.
Instead, we will investigate what happens to these shuffled averages under fluctuations over quenched disorder.
We are interested in the quenched averages of the shuffled averages:
\begin{equation}
  \begin{aligned}
    \ev{\,\ev{\Phi_\alpha(t)}_\text{sh}\,}_\mathcal{Q}
    &=
    \frac{1}{N}
    \sum_{\alpha=1}^N
    \ev{
    \Phi_\alpha(t)
    }_\mathcal{Q}
    =
    \ev{\Phi_\alpha(t)}_\mathcal{Q}
    \\
    \ev{\,\ev{\Phi_\alpha(t)\Phi_\alpha(s)}_\text{sh}\,}_\mathcal{Q}
    &=
    \frac{1}{N}
    \sum_{\alpha=1}^N
    \ev{\Phi_\alpha(t)\Phi_\alpha(s)}_\mathcal{Q}
    =
    \ev{\Phi_\alpha(t)\Phi_\alpha(s)}_\mathcal{Q}
  \end{aligned}
\end{equation}
Because the shuffled average is linear in $\Phi_\alpha(t)$ and $\ev{\Phi_\alpha(t)}_\mathcal{Q}$ is independent of $\alpha$, the quenched average of the shuffled average is just the quenched average of $\Phi_\alpha(t)$ at fixed $\alpha$.
\sidenote{
It is not a given that $\ev{\Phi_\alpha(t)}_\mathcal{Q}$ is independent of $\alpha$ in general.
In our case, it is, and you can check this but substituting the explicit expression for $\Phi_\alpha(t)$ into the average and then computing the average, just as we did in Eqs.~\eqref{eq:Phi-quenched-mean}, \eqref{eq:Phi-quenched-covariance}.
% The occurrence of these anomalous fluctuations is sometimes called replica symmetry breaking (RSB) or the breakdown of self-averaging.
In some models, there are anomalous fluctuations in which case one would find $\var_\mathcal{Q}[\ev{\Phi_\alpha(t)}_\text{sh}] = O(1)$ and proceed to compute quantities like $\ev{\ev{\Phi_\alpha(t)}_\text{sh}\ev{\Phi_\beta(t)}_\text{sh}}_\mathcal{Q}$ to understand the nature of these fluctuations. See Ref.~\cite{Nishimori} to see an example of this in the context of spin glasses.
}
The more interesting question is whether the shuffled averages fluctuate under quenched disorder.
To answer this, we can compute the variance of the shuffled averages:
% \begin{widetext}
\begin{equation}
  \begin{aligned}
    &
    \var_\mathcal{Q}
    \qty[\ev{\Phi_\alpha(t)}_\text{sh}]
    =
    \frac{1}{N^2}
    \sum_{\alpha,\beta=1}^N
    \cov_\mathcal{Q}[
      \Phi_\alpha(t),
      \Phi_\beta(t)
    ]
    \\
    &\quad=
    \frac{1}{N^2}
    \sum_{\alpha,\beta=1}^N
    \cov_\mathcal{Q}[
      \Phi_\alpha(t),
      \Phi_\beta(t)
    ]
    \\
    &\quad=
    \frac{1}{N^2}
    \sum_{\alpha,\beta=1}^N
    \sum_{i,j(\neq \alpha)}
    \sum_{k,l(\neq \beta)}
    \Bigl(
      % \\
      % &\qquad\qquad
      \ev{
      A_{\alpha i}
      A_{\beta k}
      }
      \ev{
      [
        e^{A_{\setminus \alpha} t}
      ]_{ij}
      [
        e^{A_{\setminus \beta} t}
      ]_{kl}
      }
      \ev{
      x_j(0)
      x_l(0)
      }
      \\
      &
      \qquad
      \qquad
      -
      \cancel{
      \ev{
      A_{\alpha i}
      [
        e^{A_{\setminus \alpha} t}
      ]_{ij}
      x_j(0)
      }}
      \cancel{
      \ev{
      A_{\beta k}
      [
        e^{A_{\setminus \beta} t}
      ]_{kl}
      x_l(0)
      }}
    \Bigr)
    \\
    &\quad=
    \frac{\sigma_0^2 \sigma^2}{N^3}
    \sum_{\alpha,\beta=1}^N
    \sum_{i,j(\neq \alpha)}
    \sum_{k,l(\neq \beta)}
    \qty(
      \delta_{\alpha \beta } \delta_{ik}
      -
      \delta_{\alpha k} \delta_{i\beta }
    )
    \ev{
    [
      e^{A_{\setminus \alpha} t}
    ]_{ij}
    [
      e^{A_{\setminus \beta} t}
    ]_{kl}
    }
    \delta_{jl}
    \\
    &\quad=
    \frac{\sigma_0^2 \sigma^2}{N^3}
    \Biggl(
      \sum_{\alpha}
      \sum_{i,j (\neq \alpha)}
      \ev{
      [
        e^{A_{\setminus \alpha} t}
      ]_{ij}
      [
        e^{A_{\setminus \alpha} t}
      ]_{ij}
      }
      \\
      &\qquad\qquad\qquad\qquad
      -
      \sum_{\alpha,\beta (\alpha \neq \beta)}
      \sum_{i(\neq \alpha, \beta)}
      \ev{
      [
        e^{A_{\setminus \alpha} t}
      ]_{\beta i}
      [
        e^{A_{\setminus \beta} t}
      ]_{\alpha i}
      }
    \Biggr)
    \\
    &\quad=
    \sigma_0^2 \sigma^2
    \frac{1}{N}
    \Biggl(
      2\cdot
      \frac{1}{N}
      \sum_{\alpha}
      \overbrace{
        \frac{1}{N}
        \Tr
        \qty[
        e^{A_{\setminus \alpha} t}
        \qty(e^{A_{\setminus \alpha} t})^\T
        ]
      }^{O(1)}
    \Biggr)
    \\
    &\quad=
    O\qty(
      \frac{1}{N}
    )
    .
  \end{aligned}
\end{equation}
In the final step of the calculation, we used that a trace of a random matrix like $\frac{1}{N} \Tr M$ is $O(1)$ in the thermodynamic limit $N \gg 1$ (this is because it is an average of the $O(1)$ eigenvalues of the matrix $M$).\sidenote{This is discussed in more detail in appendix~\ref{sec:brief-intro-rmt}.}
Through this computation, we've found that as $N \to \infty$, the shuffled average of $\Phi_\alpha(t)$ does not fluctuate under quenched disorder.
This means that we can replace the quenched average of the shuffled average with the quenched average of $\Phi_\alpha(t)$ at fixed $\alpha$.
In fact, this is true for all shuffled averages involving $\Phi_\alpha(t)$:
\begin{equation}
  \begin{aligned}
    \var_\mathcal{Q}
    \qty[
      \ev{
        \Phi_\alpha(t_1)
        \Phi_\alpha(t_2)
        \cdots
        \Phi_\alpha(t_n)
      }_\text{sh}
    ]
    =
    O\qty(
      \frac{1}{N}
    ).
  \end{aligned}
\end{equation}
This means that we can always replace shuffled averages of $\Phi_\alpha(t)$ with quenched averages of $\Phi_\alpha(t)$ at fixed $\alpha$:\sidenote{
  One shows this by showing that the generating functional of $\Phi_\alpha(t)$ has $O(1/N)$ variance under quenched disorder:
  \begin{align*}
    \var_\mathcal{Q}
    \qty[
      \ev{
        e^{
          \int \dd{t} \Phi_\alpha(t) j(t)
        }
      }_\text{sh}
    ]
    =
    O\qty(
      \frac{1}{N}
    )
  \end{align*}
  % We perform this calculation in appendix~\ref{sec:generating-functional-self-averaging} for the sake of completeness.
}
\begin{equation}
  \begin{aligned}
    \ev{
        \Phi_\alpha(t_1)
        \Phi_\alpha(t_2)
        \cdots
        \Phi_\alpha(t_n)
      }_\text{sh}
    &\approx
    \ev{
        \Phi_\alpha(t_1)
        \Phi_\alpha(t_2)
        \cdots
        \Phi_\alpha(t_n)
      }_\mathcal{Q},
      \\
      &\qquad\qquad\qquad\qquad\qquad\quad
    \text{for } N \gg 1.
  \end{aligned}
\end{equation}
Quantities which have this property are called \textit{self-averaging}.
For our system, this means that under frozen quenched disorder, $\Phi_\alpha(t)$ behaves like a Gaussian process where the origin of the randomness is the shuffling randomness, not the quenched disorder (which is the origin from which we first said the fluctuations came).

\subsubsection{Fluctuations of $\nu_\alpha$}

We can repeat the same sort of analysis for $\nu_\alpha(t)$.
We can compute the quenched average of $\nu_\alpha(t)$ at fixed $\alpha$:
\begin{equation}
  \begin{aligned}
    \ev{
    \nu_\alpha(t)
    }_\mathcal{Q}
		&=
		-
		\frac{1}{\sigma^2}
		\sum_{i,j(\neq \alpha)}
    \ev{
		A_{\alpha i}
    A_{j\alpha}
    }
    \ev{
		[
			e^{A_{\setminus \alpha} t}
		]_{ii}
    }
    % \\
		% &
    =
    \ev{
    \frac{1}{N}
		\sum_{i(\neq \alpha)}
		[
			e^{A_{\setminus \alpha} t}
		]_{ij}
    }
    \\
		&=
		\int \dd[2]{\lambda}
		\rho_A(\lambda)
		e^{\lambda t}
		=
		\frac{1}{2\pi\sigma^2}
		\int_{-2\sigma}^{2\sigma}
		\dd{y}
		\sqrt{4\sigma^2 - y^2}
		e^{i y t}
    \\
    &
		=
		\frac{
			J_1(2 \sigma t)
		}{
			\sigma t
		}
    .
  \end{aligned}
\end{equation}
Next, compute its fluctuations under quenched disorder:
\begin{equation}
  \begin{aligned}
    \var_\mathcal{Q}
    &
    \qty[\nu_\alpha(t)]
    =
    \frac{1}{\sigma^4}
    \sum_{i,j,k,l(\neq \alpha)}
    \cov\qty[
      A_{\alpha i}
      [
        e^{A_{\setminus \alpha} t}
      ]_{ij}
      A_{j\alpha},
      A_{\alpha k}
      [
        e^{A_{\setminus \alpha} t}
      ]_{kl}
      A_{l\alpha}
    ]
    \\
    &
    =
    \frac{1}{N^2}
    \sum_{i,j,k,l(\neq \alpha)}
    \Bigl(
    \qty(
      \delta_{ij}
      \delta_{kl}
      +
      \delta_{il}
      \delta_{jk}
      +
      \delta_{ik}
      \delta_{jl}
    )
      [
        e^{A_{\setminus \alpha} t}
      ]_{ij}
      [
        e^{A_{\setminus \alpha} t}
      ]_{kl}
    \\&\qquad\qquad\qquad\qquad\qquad\qquad
    -
    \delta_{ij}
    \delta_{kl}
      [
        e^{A_{\setminus \alpha} t}
      ]_{ij}
      [
        e^{A_{\setminus \alpha} t}
      ]_{kl}
    \Bigr)
    \\
    &
    =
    \frac{2}{N}
    \cdot
    \frac{1}{N}
    \Tr
    [
      e^{A_{\setminus \alpha} t}
      \qty(e^{A_{\setminus \alpha} t})^\T
    ]
    =
    O\qty(
      \frac{1}{N}
    )
    .
  \end{aligned}
\end{equation}
This calculation shows that $\nu_\alpha(t)$ itself (not just its average) has $O(1/N)$ fluctuations under quenched disorder.
This means that we can replace it (for all shuffling randomness and all quenched disorder) with its quenched average:
\begin{equation}
  \begin{aligned}
    \nu_\alpha(t)
    &\approx
    \ev{\nu_\alpha(t)}_\mathcal{Q}
  \end{aligned}
\end{equation}

\subsection{Putting it together: the DMFT equation of motion}

Returning to our definitions in Eqs.~\eqref{eq:dmft-eom-before-stat} and \eqref{eq:Phi-nu-definitions}, we can put everything together to write the DMFT equation of motion for the random particle $\alpha$ (at frozen quenched disorder):
\begin{equation}
	\begin{gathered}
		\dv{x}{t}
		=
		\Phi(t)
		-
		\sigma^2
		\int_0^t
		\dd{t'}
		\nu(t-t')
		x(t')
		\\
		\Phi(t)
		\sim
		\text{Gaussian Proc. (under shuffling randomness)},
    \\
		\ev{\Phi(t)}_\text{sh}
		=
		0,
		\qquad
		\ev{\Phi(t)\Phi(s)}_\text{sh}
		=
		\sigma^2
		\sigma_0^2
		\frac{J_1(2 \sigma |t-s|)}{\sigma |t-s|},
		\\
		\nu(t)
		=
		\frac{J_1(2 \sigma t)}{\sigma t}\,
		\Theta(t).
	\end{gathered}
\end{equation}
We have dropped the subscript $\alpha$ because the equation is valid for any random particle we choose.
An equation of this form is the dynamical mean-field theory (DMFT) equation of motion. In section~\ref{sec:possible-analyses}, we will discuss how this equation can provide insights into the dynamics of the system.

\section{The cavity method\label{sec:cavity-method}}

So far, we have derived the DMFT equation of motion for a single random particle in a system of $N$ particles by exactly solving the equations of motion for the other $N - 1$ particles in terms of the random particle and then substituting this solution into the equation of motion for the random particle.
An advantage of this approach was that it allowed us to identify in an exact way how the quenched disorder and shuffling randomness can be treated interchangeably for this system.
However, this approach is not very useful for non-linear systems.
There are mainly two other approaches to DMFT: the cavity method and the replica method.
In this note, we will only discuss the cavity method, which is often more direct and intuitive than the replica method.

The cavity method is a powerful technique for analyzing the dynamics of systems with quenched disorder.
It is based on the observation that order parameters like $\frac{1}{N}\sum_{i=1}^N x_i(t)$ and $\frac{1}{N}\sum_{i=1}^N x_i(t)x_i(s)$ will be nearly the same in a system of size $N$ and a system of size $N +1$ in the thermodynamic limit $N \gg 1$.
In the cavity method, one adds a new particle to the system and then analyzes how the dynamics of the system with $N+1$ particles is related to the dynamics of the system with $N$ particles.
At the end of the cavity calculation, one finds that the equation of motion for the new particle can be written in terms of the order parameters of the system with $N$ particles.
Let's see how this works.

Before we proceed, we'll write our system of equation as:
\begin{equation}
  \begin{aligned}
    \dv{x_i}{t}
    &=
    \sum_{j=1}^N
    A_{ij}
    x_j(t)
    +
    h_i(t),
  \end{aligned}
\end{equation}
where $h_i(t)$ is some auxiliary field that we can perturb to analyze the system.
At the end of the entire calculation, we will set $h_i(t) = 0$ and get the original equations of motion back.

\begin{figure}[p]
  \begin{widetext}
    \centering
    \includegraphics[]{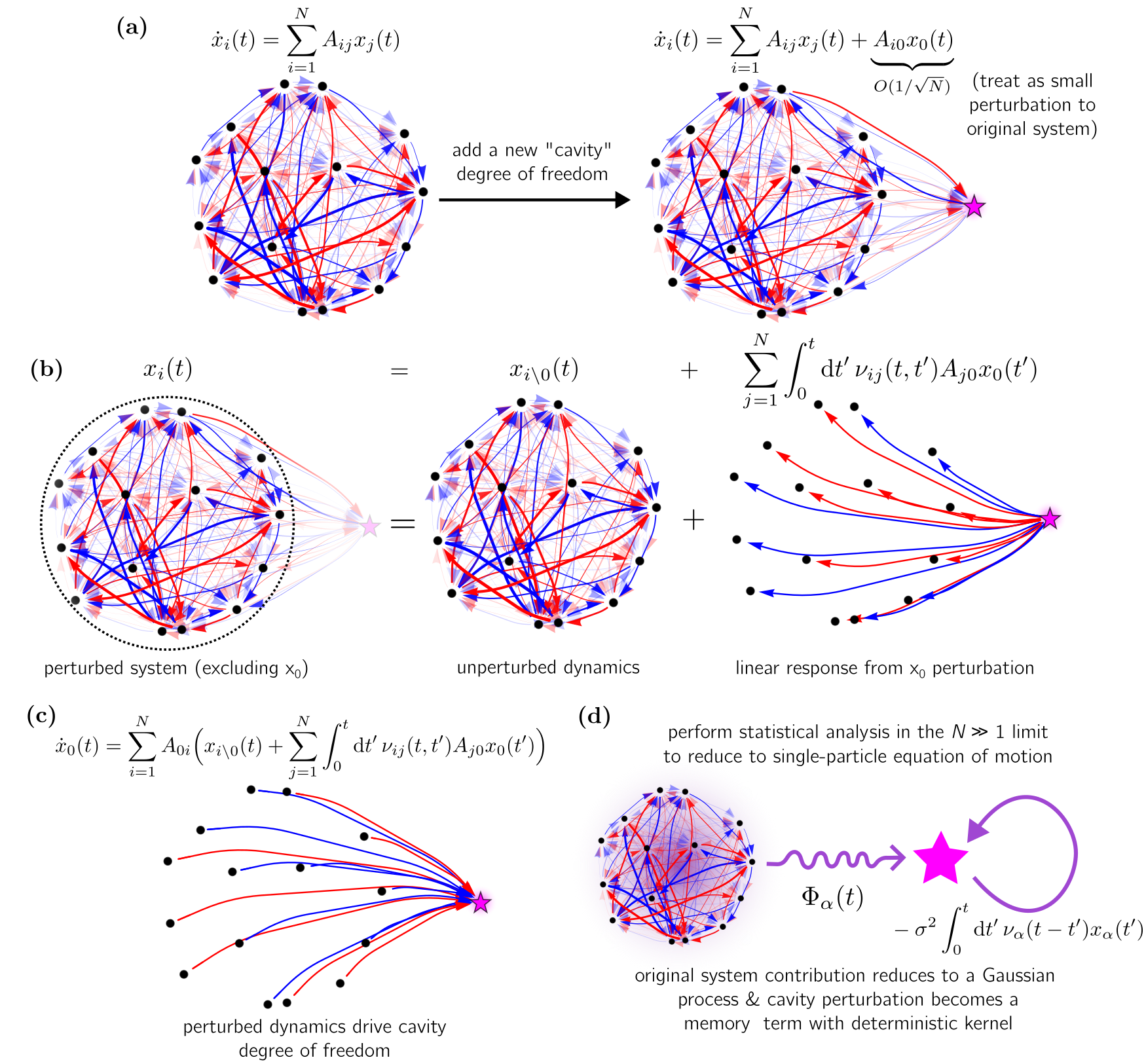}
    \caption{\label{fig:cavity-schematic}
      Schematic of the cavity method.
      (a)
      Add a new `cavity' degree of freedom $x_0(t)$ to the system which adds a new, small term, $A_{i0} x_0(t)$, to each of the equations of motion for the other particles. This is detailed in step 1 (section~\ref{sec:cavity-step-1}).
      (b)
      Treat the addition of the cavity degree of freedom as a small perturbation to the dynamics of the other particles and apply linear response theory to relate the dynamics of the system with $N + 1$ particles to the dynamics of the system with $N$ particles. 
      This results in an expression for $x_i(t)$ in terms of the dynamics before the perturbation and a memory term involving a dynamical susceptibility $\nu_{ij}(t,t')$ which describes how the dynamics of particle $i$ responds at time $t$ to a perturbation to particle $j$ at time $t'$.
      This is detailed in step 2 (section~\ref{sec:cavity-step-2}).
      (c)
      The cavity degree of freedom has dynamics driven by the dynamics of the other particles. 
      Substitute the expression for $x_i(t)$ into the equation of motion for the cavity degree of freedom to obtain an equation of motion for the cavity degree of freedom in terms of the dynamics of the original system with $N$ particles, the new coefficients $A_{0i}$ and $A_{j0}$, the dynamical susceptibility $\nu_{ij}(t,t')$, and the cavity degree of freedom itself.
      This is detailed in step 3 (section~\ref{sec:cavity-step-3}).
      (d)
      Perform a statistical analysis of the equation of motion in the thermodynamic limit $N \gg 1$ to obtain the DMFT equation of motion for the cavity degree of freedom which includes two terms: a Gaussian process which describes direct influence from the other particles and a non-local term which describes the back-reaction the cavity degree of freedom experiences after it perturbs the other particles.
      This is detailed in section~\ref{sec:cavity-step-4}.
      The final step is to `close' the DMFT by expressing the order parameters of the system in terms of the cavity degree of freedom which is discussed in section~\ref{sec:cavity-step-5}.
    }
  \end{widetext}
\end{figure}

Figure~\ref{fig:cavity-schematic} shows a schematic of the cavity method.

\subsection{Step 1: add the cavity degree of freedom\label{sec:cavity-step-1}}

The first step in the cavity method is to add a new particle to the system which is called the cavity degree of freedom and is labeled $i = 0$.
The new equations of motion for the system with $N + 1$ particles are:
\begin{equation}
  \begin{aligned}
    \dv{x_i}{t}
    &=
    \sum_{j=1}^N
    A_{ij}
    x_j(t)
    +
    A_{i0}
    x_0(t)
    +
    h_i(t),
    \quad
    i = 1, \ldots, N,
    \\
    \dv{x_0}{t}
    &=
    \sum_{i=1}^N
    A_{0i}
    x_i(t)
    +
    A_{00}
    x_0(t)
    +
    h_0(t).
  \end{aligned}
\end{equation}
Notice that we can drop the second term in the second equation because $A_{ij}$ is skew-symmetric.
In more general contexts, this term might not be zero, but we still drop it because it is of order $1/\sqrt{N}$.

\subsection{Step 2: treat the cavity degree of freedom as a perturbation\label{sec:cavity-step-2}}

Notice that for particles $i = 1,\ldots,N$, the equations of motion are the same as before except for the new term $A_{i0} x_0(t)$.
We can treat this as an effective perturbation to the auxiliary field $h_i(t)$:
\begin{equation}
  \begin{aligned}
    h_i(t)
    \rightsquigarrow 
    h_i(t) + A_{i0} x_0(t).
  \end{aligned}
\end{equation}
Because $A_{i0}$ is $O(1/\sqrt{N})$, we can treat this as a small perturbation and use perturbation theory to relate the dynamics of the system with $N + 1$ particles to the dynamics of the system with $N$ particles.
The linear response approximation is:
\sidenote{
Here, $\fdv{F[h(t)]}{h(t')}$ is the functional derivative of the functional $F[h(t)]$ with respect to the field $h(t')$.
It measures how the function $F[h(t)]$ changes to linear order when we perturb the field $h(t)$ at time $t'$.
It is defined so that $\fdv{h(t)}{h(t')} = \delta(t - t')$ where $\delta$ is the Dirac delta function.
}
\begin{equation}
  \begin{gathered}
    x_i(t)
    =
    x_{i\setminus0}(t)
    +
    \sum_{j=1}^N
    \int_0^t
    \dd{t'}
    \nu_{ij}(t,t')
    A_{j0}
    x_0(t')
    +
    O(
      N^{-1/2}
    )
    ,
    \\
    \text{where:}\qquad
    \nu_{ij}(t,t')
    \equiv
    \eval{
      \fdv{x_i(t)}{h_j(t')}
    }_{h=0}
    .
  \end{gathered}
\end{equation}

\subsection{Step 3: obtain the cavity equation of motion\label{sec:cavity-step-3}}

Now, we can substitute this expression for $x_i(t)$ into the equation of motion for the cavity degree of freedom:
\begin{equation}
  \begin{aligned}
    \dv{x_0}{t}
    &=
    \sum_{i=1}^N
    A_{0i}
    x_{i\setminus0}(t)
    +
    \sum_{i,j=1}^N
    \int_0^t
    \dd{t'}
    \nu_{ij}(t,t')
    A_{0i}
    A_{j0}
    x_0(t')
    +
    h_0(t)
    .
  \end{aligned}
  \label{eq:cavity-eom-before-stat}
\end{equation}
What's notable about this equation is that it is an equation for $x_0(t)$ purely in terms of the dynamics of the system with $N$ particles, the new coefficients $A_{0i}$ and $A_{j0}$, and $x_0$ itself.
\sidenote{
Note that up to this point, everything we have done is exact as $N \to \infty$. The only approximation we have made is that we have treated the new term $A_{i0} x_0(t)$ as a small perturbation, but the error is of order $O(1/\sqrt{N})$ which is negligible in the thermodynamic limit $N \gg 1$.
}
To keep things tidy, we define:
\begin{equation}
  \begin{aligned}
    \Phi_0(t)
    \equiv
    \sum_{i=1}^N
    A_{0i}
    x_{i\setminus0}(t).
  \end{aligned}
\end{equation}

\subsection{Step 4: statistics under quenched disorder\label{sec:cavity-step-4}}

So far, we have assumed that the coefficients $A_{ij}$ are fixed and do not fluctuate.
Now, let's consider how each term in Eq.~\eqref{eq:cavity-eom-before-stat} fluctuates under quenched disorder.
Notice that the first term, $\Phi_0(t)$, is a sum of many normal random variables $\{A_{0i}\}$, so it is a Gaussian process.
This means that $\Phi_0(t)$ is fully characterized by its mean and covariance:
\begin{gather}
  \begin{aligned}
    \ev{
      \Phi_0(t)
    }_\mathcal{Q}
    =
    \sum_{i=1}^N
    \ev{A_{0i}}
    \cancel{\ev{x_{i\setminus0}(t)}}
    =
    0
  \end{aligned}
  \\
  \begin{aligned}
    \ev{
      \Phi_0(t)
      \Phi_0(s)
    }_\mathcal{Q}
    &=
    \sum_{i,j=1}^N
    \ev{A_{0i} A_{0j}}
    x_{i\setminus0}(t)
    x_{j\setminus0}(s)
    \\
    &=
    \frac{\sigma^2}{N}
    \sum_{i,j=1}^N
    \delta_{ij}
    x_{i\setminus0}(t)
    x_{j\setminus0}(s)
    =
    \sigma^2\,
    C(t,s),
  \end{aligned}
\end{gather}
where we have defined:
\begin{equation}
  \begin{aligned}
    C(t,s)
    \equiv
    \frac{1}{N}
    \sum_{i=1}^N
    x_{i\setminus0}(t)
    x_{i\setminus0}(s),
  \end{aligned}
\end{equation}
to be the empirical correlation function.
As for the second term:
\begin{equation}
  \begin{aligned}
    \ev{
      \sum_{i,j=1}^N
      \nu_{ij}(t,t')
      A_{0i}
      A_{j0}
    }
    % &\qquad
    &=
    \sum_{i,j=1}^N
    \nu_{ij}(t,t')
    \ev{
      A_{0i}
      A_{j0}
    }
    \\
    &
    =
    -
    \frac{\sigma^2}{N}
    \sum_{i,j=1}^N
    \delta_{ij}\,
    \nu_{ij}(t,t')
    =
    -
    \sigma^2
    \nu(t,t')
    ,
  \end{aligned}
\end{equation}
where we have defined:
\begin{equation}
  \begin{aligned}
    \nu(t,t')
    \equiv
    \frac{1}{N}
    \sum_{i=1}^N
    \nu_{ii}(t,t')
    =
    \frac{1}{N}
    \sum_{i=1}^N
    \eval{
      \fdv{x_i(t)}{h_i(t')}
    }_{h=0}
    ,
  \end{aligned}
\end{equation}
to be the average dynamical susceptibility.
The variance is:\sidenote{
Here we used Wick's theorem to compute the expectation of the product of four Gaussian random variables.
This is discussed in more detail in appendix section~\ref{sec:wick-theorem}.
}
\begin{equation}
  \begin{aligned}
    &
    \var_\mathcal{Q}
    \qty[
    \sum_{i,j=1}^N
    \nu_{ij}(t,t')
    A_{0i}
    A_{j0}
    ]
    =
    \sum_{i,j,k,l=1}^N
    \cov
    \qty[
    A_{0i}
    A_{j0}
    ,
    A_{0k}
    A_{l0}
    ]
    \nu_{ij}(t,t')
    \nu_{kl}(t,t')
    \\
    &
    \qquad
    =
    \frac{\sigma^2}{N^2}
    \sum_{i,j,k,l=1}^N
    \qty(
      \delta_{ik}
      \delta_{jl}
      +
      \delta_{il}
      \delta_{jk}
    )
    \nu_{ij}(t,t')
    \nu_{kl}(t,t')
    \\
    &
    \qquad
    =
    \frac{\sigma^2}{N}
    \Biggl(
      \underbrace{
      \frac{1}{N}
      \sum_{i,j=1}^N
      \nu_{ij}(t,t')
      \nu_{ij}(t,t')
      }_{O(1)}
      +
      \underbrace{
      \frac{1}{N}
      \sum_{i,j=1}^N
      \nu_{ij}(t,t')
      \nu_{ji}(t,t')
      }_{O(1)}
    \Biggr)
    \\
    &
    \qquad
    =
    O\qty(
      \frac{1}{N}
    )
    .
  \end{aligned}
\end{equation}
In the final step of the calculation, we again used that the scaled trace of a random matrix like $\frac{1}{N} \Tr M$ is $O(1)$ in the thermodynamic limit $N \gg 1$.
This result means that the second term in Eq.~\eqref{eq:cavity-eom-before-stat} has $O(1/N)$ fluctuations under quenched disorder and can thus be replaced with its quenched average:
\begin{equation}
  \begin{aligned}
    \sum_{i,j=1}^N
    \nu_{ij}(t,t')
    A_{0i}
    A_{j0}
    &\approx
    -
    \sigma^2
    \nu(t,t').
  \end{aligned}
\end{equation}
Putting this together, we reduce the cavity equation of motion in Eq.~\eqref{eq:cavity-eom-before-stat} to effective dynamics (where fluctuations are due to quenched disorder):
\begin{equation}
  \label{eq:dmft-eom-cavity}
  \begin{gathered}
    \dv{x_0}{t}
    =
    \Phi_0(t)
    -
    \sigma^2
    \int_0^t
    \dd{t'}
    \nu(t,t')
    x_0(t')
    +
    h_0(t)
    \\
    \Phi_0(t)
    \sim
    \text{Gaussian Proc. (under quenched disorder)},
    \\
    \ev{\Phi_0(t)}_\mathcal{Q}
    =
    0,
    \quad
    \ev{\Phi_0(t)\Phi_0(s)}_\mathcal{Q}
    =
    \sigma^2
    C(t,s)
    =
    \sigma^2
    \frac{1}{N}
    \sum_{i=1}^N
    x_{i\setminus0}(t)
    x_{i\setminus0}(s),
    \\
    \nu(t,t')
    =
    \frac{1}{N}
    \sum_{i=1}^N
    \eval{
      \fdv{x_i(t)}{h_i(t')}
    }_{h=0}
    .
  \end{gathered}
\end{equation}

\subsection{Step 5: assert self-averaging to get self-consistent equations\label{sec:cavity-step-5}}

We now have an equation of motion for the cavity degree of freedom $x_0(t)$ in terms of the order parameters $C(t,s)$ and $\nu(t,t')$.
However, based on the DMFT equation alone, we have no way to compute these order parameters.
To compute them, we need to make an assumption about the self-averaging of the system.
In particular, we assert that for the order parameters, their quenched averages are equal to their shuffled averages as $N \to \infty$:
\begin{equation}
  \begin{gathered}
    C(t,s)
    =
    \overbrace{
    \frac{1}{N}
    \sum_{i=1}^N
    x_{i\setminus0}(t)
    x_{i\setminus0}(s)
    }^\text{``shuffled average''}
    =
    \overbrace{\ev{x_0(t)x_0(s)}_\mathcal{Q}}^\text{``quenched average''}
    \\
    \nu(t,t')
    =
    \frac{1}{N}
    \sum_{i=1}^N
    \eval{
      \fdv{x_i(t)}{h_i(t')}
    }_{h=0}
    =
    \ev{
      \eval{
        \fdv{x_0(t)}{h_0(t')}
      }_{h=0}
    }_\mathcal{Q}
  \end{gathered}
\end{equation}
These sorts of expressions are called \textit{closure} or \textit{self-consistency} conditions for the DMFT.
\sidenote{
Another order parameter,
\begin{align*}
  m(t)
  \equiv
  \frac{1}{N}
  \sum_{i=1}^N
  x_{i\setminus0}(t),
\end{align*}
often appears in DMFT calculations, but it did not show up here because it is zero for our model.
The closure/self-consistency condition for $m(t)$ is:
\begin{align*}
  m(t)
  =
  \ev{x_0(t)}_\mathcal{Q}
  .
\end{align*}
}
We showed that these self-averaging properties are true for our model in the previous section by explicitly computing the fluctuations of the order parameters under quenched disorder.
However, this is not possible for most systems, so the above `closing' of the DMFT equations is usually done by \textit{assuming} that the order parameters are self-averaging.

\subsection{Step 6: solve the self-consistent equations (hard!)}

We'll first solve for the dynamical susceptibility $\nu(t,t')$.
We do this by taking a functional derivative of the DMFT equation of motion with respect to $h_0(t')$:
\begin{equation}
  \begin{aligned}
    \pdv{t}
    \nu(t,t')
    &
    =
    \pdv{t}
    \ev{
      \fdv{x_0(t)}{h_0(t')}
    }
    =
    \ev{
      \fdv{h_0(t')}
      \dv{x_0(t)}{t}
    }
    \\
    &
    =
    -
    \sigma^2
    \int_0^t
    \dd{t''}
    \nu(t,t'')
    \nu(t'',t')
    +
    \delta(t - t')
    .
  \end{aligned}
  \label{eq:dmft-eom-cavity-nu}
\end{equation}
This along with the initial condition $\nu(t,t') = 0$ for $t < t'$ is a self-consistent equation for $\nu(t,t')$.
This equation implies that $\nu(t,t')$ is a function of the difference $t - t'$ alone, so we can write $\nu(t,t') = \nu(t - t')$.
Let:
\begin{equation}
  \begin{aligned}
    \widetilde{\nu}(z)
    \equiv
    \int_0^\infty
    \dd{t}
    \nu(t)
    e^{-z t},
  \end{aligned}
\end{equation}
be the Laplace transform of $\nu(t)$.
Then Eq.~\eqref{eq:dmft-eom-cavity-nu} becomes:
\begin{equation}
  \begin{aligned}
    z \widetilde{\nu}(z)
    &=
    -
    \sigma^2
    [\widetilde{\nu}(z)]^2
    +
    1
    \implies
    \widetilde{\nu}(z)
    =
    % \frac{
    %   2
    % }{
    %   z + \sqrt{z^2 + 4 \sigma^2}
    % }
    \frac{1}{2\sigma^2}
    \qty(
      \sqrt{
        z^2 + 4\sigma^2
      }
      -
      z
    )
   \end{aligned}
\end{equation}
The inverse Laplace transform of this expression gives:
\begin{equation}
  \begin{aligned}
    \nu(t)
    =
    \frac{J_1(2 \sigma t)}{\sigma t}
    \, \Theta(t),
  \end{aligned}
\end{equation}
where $J_1(x)$ is the Bessel function of the first kind of order 1 and $\Theta(t)$ is the Heaviside step function (which is 1 for $t > 0$ and 0 for $t < 0$).
\sidenote{
There are a few ways to compute this inverse Laplace transform, but the easiest way is to look at a table of Laplace transforms (on Wikipedia, for example) or use computer algebra software like Mathematica.
}
Now that we have $\nu(t)$, we can obtain an explicit expression for $x_0(t)$ in terms of $\Phi(t)$.
If we write:
\begin{equation}
  \begin{aligned}
    x_0(t)
    =
    G(t)
    x_0(0)
    +
    \int_0^t
    \dd{t'}
    G(t - t')
    \Phi_0(t'),
  \end{aligned}
\end{equation}
and substitute this back into the Eq.~\eqref{eq:dmft-eom-cavity}, we find that this is a general solution to the DMFT equation of motion when:
\begin{equation}
  \begin{aligned}
    \dv{G}{t}
    &=
    \delta(t)
    -
    \sigma^2
    \int_0^t
    \dd{t'}
    \nu(t - t')
    G(t')
  \end{aligned}
\end{equation}
with initial condition $G(t) = 0$ for $t < 0$.
This equation can be solved using the Laplace transform:
\begin{align}
  \label{eq:dmft-eom-cavity-G-resolvant-equation}
  z \widetilde{G}(z)
  =
  &
  1
  -
  \sigma^2
  \widetilde{\nu}(z)
  \widetilde{G}(z)
  \\
  &
  \implies
  \widetilde{G}(z)
  =
  \frac{1}{z + \sigma^2 \widetilde{\nu}(z)}
  =
  \frac{1}{2\sigma^2}
  \qty(
    \sqrt{
      z^2 + 4\sigma^2
    }
    -
    z
  )
\end{align}
which we have already seen has the inverse Laplace transform:
\sidenote{
The fact that $G(t) = \nu(t)$ appears in the solution is special to this model and is not true in general.
% It is a consequence of the fact that the matrix $A_{ij}$ is skew-symmetric and has a simple spectrum.
}
\begin{equation}
  \begin{aligned}
    G(t)
    =
    \frac{J_1(2 \sigma t)}{\sigma t}
    \, \Theta(t).
  \end{aligned}
\end{equation}
To close the DMFT equations, we compute:
\begin{equation}
  \begin{aligned}
    C(t,s)
    &
    =
    \ev{
      x_0(t)
      x_0(s)
    }
    \\
    &
    =
    \sigma_0^2
    G(t)G(s)
    % \\
    % &\qquad
    +
    \sigma^2
    \int_0^t
    \dd{t'}
    \int_0^s
    \dd{s'}
    G(t - t')
    G(s - s')
    C(t',s')
  \end{aligned}
\end{equation}
If we define the two-variable Laplace transform:
\begin{equation}
  \begin{aligned}
    \widetilde C(z_t, z_s)
    =
    \int_0^\infty
    \dd{t}
    \int_0^\infty
    \dd{s}
    e^{-z_t t - z_s s}
    C(t,s),
  \end{aligned}
\end{equation}
we can apply the convolution theorem to write:
\begin{equation}
  \begin{aligned}
    \widetilde{C}(z_t, z_s)
    &=
    \sigma_0^2
    \widetilde{G}(z_t) \widetilde{G}(z_s)
    +
    \sigma^2
    \widetilde{G}(z_t) \widetilde{G}(z_s)
    \widetilde{C}(z_t, z_s)
  \end{aligned}
\end{equation}
which can be rearranged to give:
\begin{equation}
  \begin{aligned}
    \widetilde{C}(z_t, z_s)
    &=
    \frac{
      \sigma_0^2 \widetilde{G}(z_t) \widetilde{G}(z_s)
    }{
      1 - \sigma^2 \widetilde{G}(z_t) \widetilde{G}(z_s)
    }
  \end{aligned}
\end{equation}
Using Eq.~\eqref{eq:dmft-eom-cavity-G-resolvant-equation}, we can write this as:
\begin{equation}
  \begin{aligned}
    \widetilde{C}(z_t, z_s)
    =
    \sigma_0^2 
    \frac{
      \widetilde{G}(z_t) \widetilde{G}(z_s)
    }{
      z_t + z_s
    }
  \end{aligned}
\end{equation}
In general:
\begin{equation}
  \begin{aligned}
    \int_0^\infty
    \dd{t}
    \int_0^\infty
    \dd{s}
    e^{-z_t t - z_s s}
    f(t-s)
    &=
    \frac{
      \widetilde{f}(z_t)
      \widetilde{f}(z_s)
    }{
      z_t + z_s
    }
  \end{aligned}
\end{equation}
for any well-behaved function $f(t)$ which tells us that $C(t,s)$ is a function of the difference $t - s$ alone and that:
\begin{equation}
  \begin{aligned}
    C(t-s)
    &=
    \sigma_0^2
    G(|t-s|)
    =
    \sigma_0^2
    \frac{J_1(2 \sigma |t-s|)}{\sigma |t-s|}
    .
  \end{aligned}
\end{equation}
\begin{marginfigure}%[10mm]
  \includegraphics[width=\marginparwidth]{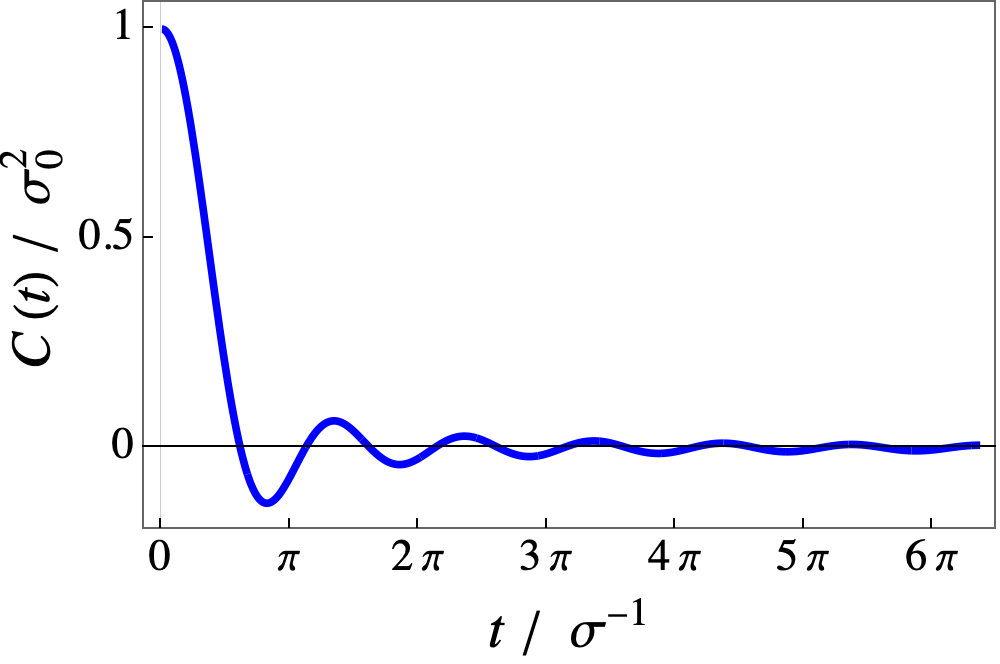}
  \caption{\label{fig:correlation-function-plot}
  Plot of the correlation function $C(t-s) = \ev{x(t)x(s)}$. The oscillations have period $\pi/\sigma$.
  Because $C(t) = \sigma_0^2 G(t) = \sigma_0^2 \nu(t)$, this plot also shows the shape of the dynamical susceptibility $\nu(t)$ and the Green's function $G(t)$.
  }
\end{marginfigure}

In the long time limit:
\begin{equation}
  \begin{aligned}
    C(t)
    \sim
    \frac{
      \sigma_0^2
    }{
      \sqrt{\pi}
      \sigma^{3/2}
      t^{3/2}
    }
    \sin
    \qty(
      2 \sigma t
      -
      \frac{\pi}{4}
    ),
    \quad
    \text{for } t \gg 1/\sigma.
  \end{aligned}
\end{equation}
This means that there are coherent oscillations in $x(t)$ which decay slowly as $t^{-3/2}$.
Putting this all together Eq.~\eqref{eq:dmft-eom-cavity} becomes:
\begin{equation}
  \label{eq:dmft-eom-cavity-solved}
  \begin{gathered}
    \dv{x}{t}
    =
    \Phi(t)
    -
    \sigma^2
    \int_0^t
    \dd{t'}
    \nu(t,t')
    x(t')
    \\
    \Phi(t)
    \sim
    \text{Gaussian Proc. (under shuffling disorder)},
    \\
    \ev{\Phi(t)}_\mathcal{Q}
    =
    0,
    \quad
    \ev{\Phi(t)\Phi(s)}_\mathcal{Q}
    =
    \sigma^2
    \sigma_0^2
    \frac{J_1(2 \sigma |t-s|)}{\sigma |t-s|},
    \\
    \nu(t,t')
    =
    \frac{1}{N}
    \sum_{i=1}^N
    \eval{
      \fdv{x_i(t)}{h_i(t')}
    }_{h=0}
    =
    \frac{J_1(2 \sigma |t-t'|)}{\sigma |t-t'|}\,
    \Theta(t-t')
    ,
  \end{gathered}
\end{equation}
which precisely matches the DMFT equation of motion we derived in section~\ref{sec:exactly-deriving-dmft-linear-model}.

\subsection{Why was it necessary to add the cavity degree of freedom?}

When deriving the DMFT equation of motion, we might have been tempted to look at the sum $\sum_{j(\neq i)} A_{ij} x_j(t)$, recognize that it is the sum of many normal random variables, and conclude that it is a Gaussian process with mean zero and variance $\sigma^2 C(t,s)$.
Knowing what we know now, we can see that this would miss the back-reaction term, $-\sigma^2 \int_0^t \dd{t'} \nu(t,t') x_i(t')$, which is crucial for the DMFT.
We miss this back-reaction term because $x_i(t)$ has $O(1/\sqrt{N})$ contributions to the dynamics of all other degrees of freedom $j\neq i$ which each then have $O(1/\sqrt{N})$ contributions to the dynamics of $x_i(t)$ which all add up to give an $O(1)$ contribution to the dynamics of $x_i(t)$.
This feedback is not captured by simply taking averages of $\sum_{j(\neq i)} A_{ij} x_j(t)$ because we essentially assumed that $x_j$ is independent of $A_{ij}$, which would be the case if $x_j(t)$ were a deterministic function, but we should actually treat it as a stochastic process that is correlated with $A_{ij}$.
When we add the cavity degree of freedom, we instead get the sum: $\Phi_0(t) = \sum_{i=1}^N A_{0i} x_{i\setminus0}(t)$.
Because $A_{0i}$ and $x_{i\setminus0}(t)$ are independent (by definition), we are free to treat averages over $A_{0i}$ and $x_{i\setminus0}(t)$ separately.

\section{A taste of possible analyses\label{sec:possible-analyses}}

Now that we have derived the DMFT equation of motion, let's explore how it might be used to analyze the dynamics of the system.

\begin{marginfigure}
  \includegraphics[width=\marginparwidth]{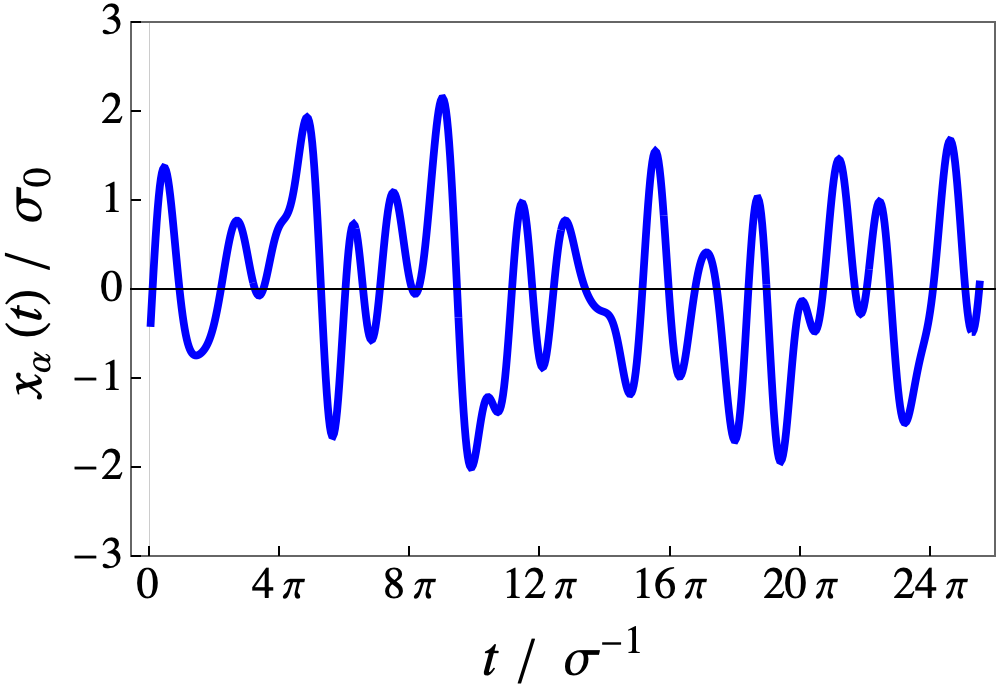}
  \caption{\label{fig:example-dynamics-plot}
  % \textbf{WRITE CAPTION} N = 64
  Plot of the dynamics of a single degree of freedom $x_\alpha(t)$ in a system $N = 64$.
  }
\end{marginfigure}

\subsection{Correlations}

\begin{marginfigure}
  \includegraphics[width=\marginparwidth]{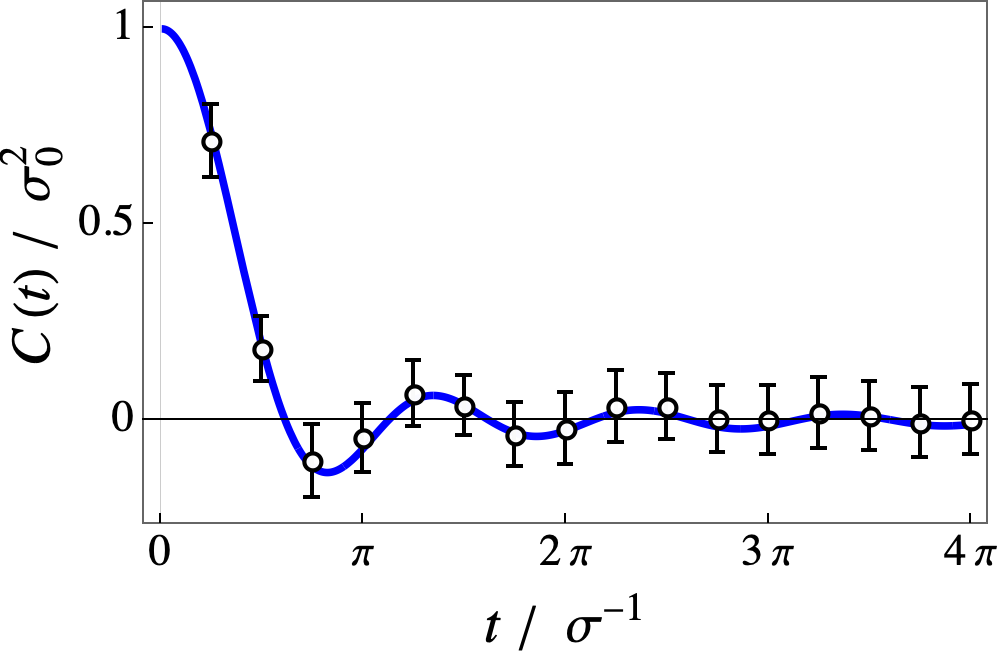}
  \caption{\label{fig:correlation-function-plot-with-dat}
  Plot of the covariance function $C(t_0,t_0 + t) = \frac{1}{N}\sum_{i=1}^N x_i(t_0) x_i(t_0 + t)$ from the DMFT prediction (solid line) and from numerical simulations (dots).
  The numerical simulations were done with $N = 128$ and averaged over $32$ random realizations of the dynamics.
  The error bars are the standard deviation of $C(t_0,t_0 + t)$ from replicate to replicate.
  }
\end{marginfigure}

The most immediate thing to ask is: if I look at the state of a random degree of freedom $x_i(t)$ at time $t$, how correlated is it with its own state at some later time $t+\tau$?
This question is only well-defined in the case where we are looking at a random degree and integrating out the other degrees of freedom, otherwise everything is deterministic and perfectly correlated with itself.
In solving the self-consistency equations in the previous section, we already determined $C(t,s) = \ev{x_i(t)x_i(s)} = \frac{1}{N}\sum_{i=1}^N x_i(t)x_i(s)$.
We can then determine the correlations:
\begin{equation}
  \begin{aligned}
    \corr[x_i(t), x_i(t+\tau)]
    &=
    \frac{
      \ev{x_i(t)x_i(t+\tau)}
      -
      \ev{x_i(t)}\ev{x_i(t+\tau)}
    }{
      \sqrt{
        \var[x_i(t)]
        \var[x_i(t+\tau)]
      }
    }
    \\
    &
    =
    \frac{C(t,t+\tau)}{\sigma_0^2}
    =
    \frac{J_1(2 \sigma \tau)}{\sigma \tau}
    .
  \end{aligned}
\end{equation}
The shape of this function is shown in figure~\ref{fig:correlation-function-plot-with-dat} along with a comparison to numerical simulations.
We can see that there are oscillations between positive and negative correlation with approximate period $\pi/\sigma$ which decay slowly as $\tau^{-3/2}$.
This is remarkable because it says that there is a quasi-periodic structure to the dynamics of each degree of freedom which is coherent for a long time.
This quasi-periodic structure can be seen in figure~\ref{fig:example-dynamics-plot}.
Ultimately, this is a consequence of the fact that the matrix $A_{ij}$ is skew-symmetric, meaning that its eigenvalues are purely imaginary, corresponding to oscillatory dynamics.

\subsection{Response to perturbations}

Another interesting question is: if I perturb a random degree of freedom $x_i$ at time $t_0$ by a small amount, how much does it affect its own state at some later time $t_0 + t$?
Because our equation is linear, we can interpret this as adding a small perturbation $h_i(t) = \epsilon \delta(t - t_0)$ to the right-hand-side of the equation of motion for $x_i(t)$.
This response to perturbations is then captured by the dynamical susceptibility $\fdv*{x_i(t_0 + t)}{h_i(t_0)}$.
Our result was that the average dynamical susceptibility is:
\begin{equation}
  \begin{aligned}
    \nu(t_0+t,t_0)
    =
    \frac{1}{N}
    \sum_{i=1}^N
    \eval{
      \fdv{x_i(t_0 + t)}{h_i(t_0)}
    }_{h=0}
    =
    \frac{J_1(2 \sigma t)}{\sigma t}
    .
  \end{aligned}
\end{equation}
This functional form is identical to the correlation function, so you can see a plot of $\nu(t)$ in Fig.~\ref{fig:correlation-function-plot}.
This tells us that whether the effect of a perturbation is positive or negative depends on the time delay $t$ in a quasi-periodic way and that the effect of the perturbation decays slowly as $t^{-3/2}$.

A common quantity of interest, especially in non-linear systems, is the Lyapunov exponent, which measures how quickly two nearby trajectories diverge from one another.
Assume we perturb the system at time $t_0$ by some small amount $\epsilon$ in a direction $\vb*\xi$ so that: $\vb*x'(t_0) = \vb*x(t_0) + \epsilon \,\vb*\xi$.
The Lyapunov exponent is defined as:
\begin{equation}
  \begin{aligned}
    \lambda
    =
    \lim_{\epsilon \to 0}
    \lim_{t \to \infty}
    \frac{1}{t}
    \ln
    \frac{
      \norm{\vb*x'(t_0 + t) - \vb*x(t_0 + t)}
    }{
      \epsilon
    }
  \end{aligned}
\end{equation}
In the case where $\xi_i \sim \mathcal{N}(0,1)$ are i.i.d. normal random variables, we can write the average Lyapunov exponent as:
\sidenote{
Taking the average over $\vb*\xi$ is necessary mainly for formal mathematical reasons, but usually does not matter in practice because the Lyapunov exponent is self-averaging over $\vb*\xi$.
Additionally, this sort of Lyapunov exponent ultimately converges numerically to the same value for almost all choices of $\vb*\xi$, so long as $\vb*\xi$ is chosen `generically' (i.e. not in a special direction).
}
\begin{equation}
  \begin{aligned}
    \ev{\lambda}_\xi
    =
    \lim_{t_0,t \to \infty}
    \frac{1}{2t}
    \ln
    \frac{1}{N}
    \sum_{i,j = 1}^N
    \nu_{ij}(t_0+t, t)^2,
  \end{aligned}
\end{equation}
where the average is taken over the random perturbation direction $\vb*\xi$ \cite{SompolinskyPathIntegral}.
The quantity $\frac{1}{N}
\sum_{i,j = 1}^N
\nu_{ij}(t_0+t, t)^2$ is self-averaging over quenched disorder and also happens to be the second moment of the eigenvalue spectrum of $\nu_{ij}(t_0+t, t)$.
To determine this quantity, one performs a two-particle cavity calculation in which one introduces two cavity degrees of freedom.
This quantity has historically played an important role in the study of chaotic dynamics in neural networks \cite{SompolinskyPRL,SompolinskyPathIntegral,SompolinskyX,MoritzPRX}.
In our simple linear model, this quantity is exactly zero for all choices of parameters because the eigenvalues of $A_{ij}$ are purely imaginary.

\subsection{Fourier modes and power spectrum}

We can also ask: if I look at the Fourier modes of a random degree of freedom $x_i(t)$, how much power is in each mode?
This measures how much of the dynamics is happening at different frequencies.
The power spectral density of $x_i(t)$ at frequency $\omega$ is defined as:
\begin{equation}
  \begin{aligned}
    S_x(\omega)
    =
    \ev{
      \lim_{T \to \infty}
      \frac{1}{T}
      \abs{
        \int_0^T
        \dd{t}
        x_i(t)
        e^{-i \omega t}
      }^2
    }.
  \end{aligned}
\end{equation}
It is the average total overlap of $x_i(t)$ with the Fourier mode $e^{i \omega t}$ per unit time.
Because the DMFT for $x_i(t)$ is a Gaussian process\sidenote{
Note that this sort of analysis might not work for all models because the application of the Wiener--Khinchin theorem requires that the process is stationary, ergodic, and Gaussian which is often not the case for non-linear models.
}, we can compute the power spectral density using the Wiener--Khinchin theorem:
\begin{equation}
  \begin{aligned}
    S_x(\omega)
    &=
    % \frac{1}{\sqrt{2\pi}}
    \int_{-\infty}^\infty
    \dd{t}
    C(t)
    e^{-i \omega t}
    =
    2
    \sigma_0^2
    \int_0^\infty
    \dd{t}
    \frac{J_1(2 \sigma t)}{\sigma t}
    e^{-i \omega t}
  \end{aligned}
\end{equation}
To evaluate this integral, we use the Laplace identity:
\begin{equation}
  \begin{aligned}
    \mathcal{L}_t\qty[
      \frac{J_1(2\sigma t)}{\sigma t}
    ](z)
    &=
    \frac{1}{2\sigma^2}
    \qty(
      \sqrt{z^2 + 4\sigma^2}
      -
      z
    )
    .
  \end{aligned}
\end{equation}
However, we are working in the complex plane, so we need to be careful about branch cuts of the square root function.
Choosing the branch cut that ensures $S_x(\omega) \geq 0$ for all $\omega$, we find:
\begin{marginfigure}
  \includegraphics[width=\marginparwidth]{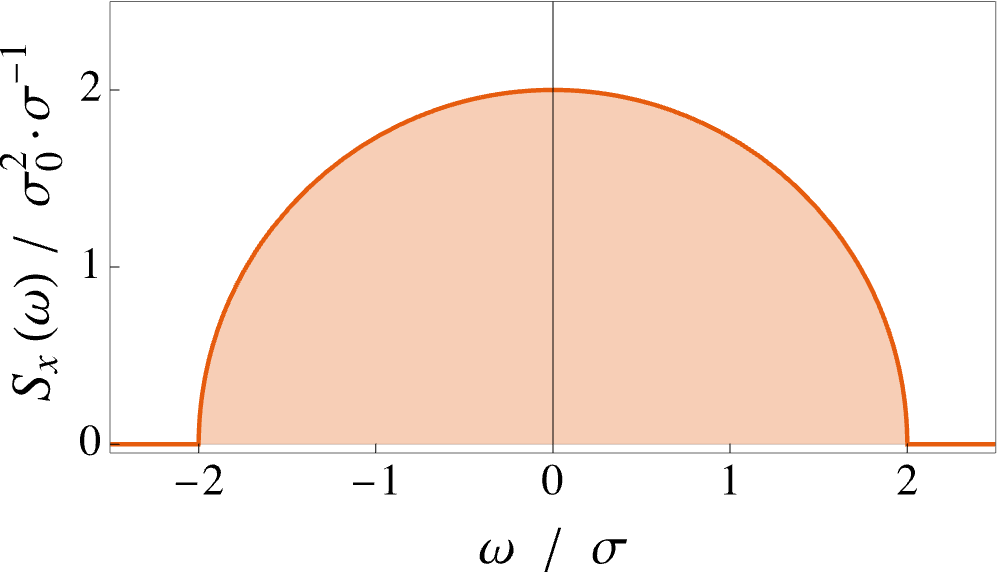}
  \caption{\label{fig:power-spec-dens}
    Power spectral density $S_x(\omega)$ of the dynamics of a random degree of freedom $x_i(t)$ in the DMFT model.
    This quantity is the average total overlap of $x_i(t)$ with the Fourier mode $e^{i \omega t}$ per unit time.
  }
\end{marginfigure}
\begin{equation}
  \begin{aligned}
    S_x(\omega)
    =
    \begin{cases}
      \frac{\sigma_0^2}{\sigma^2}
      \sqrt{4\sigma^2 - \omega^2},
      & \abs{\omega} < 2\sigma
      \\
      0,
      & \abs{\omega} > 2\sigma
    \end{cases}
    .
  \end{aligned}
\end{equation}
This is not too surprising because this is the same shape as the Wigner semicircle distribution, which is the eigenvalue spectrum of $A_{ij}$.
This result means that the dynamics of $x_i(t)$ are dominated by frequencies in the range $[-2\sigma, 2\sigma]$ and that there is a sharp cutoff in power at these frequencies.
Additionally, this means that if we were to drive the system with an external oscillatory input at frequency $\omega$, the system would only respond if $\abs{\omega} < 2\sigma$.
Because the power spectral density peaks at $\omega = 0$ and increases as $\abs{\omega} \to 0$, the system is most responsive to low-frequency inputs and has more slow dynamics than fast dynamics.

\section{A non-linear example: the generalized Lotka--Volterra model\label{sec:generalized-lotka-volterra}}

Finally, we briefly discuss how the cavity method can be applied to a non-linear model, specifically the generalized Lotka--Volterra (GLV) model of interacting species.
This derivation roughly follows Ref.~\cite{Roy_2019}.

\subsection{The model}

The GLV model describes the populations $N_i$ of species $i = 1,\ldots, S$.
It is described by the coupled ordinary differential equations: \cite{bunin}
\begin{equation}
  \begin{aligned}
    \dv{N_i}{t}
    =
    N_i
    \qty(
      1 
      -
      N_i
      -
      \sum_{j(\neq i)}
      A_{ij}
      N_j
    )
    ,\qquad
    i = 1,\ldots,S
    .
  \end{aligned}
\end{equation}
This is a phenomenological model of species interactions in which a species $i$ grows logistically in isolation, but in the presence of another species $j$, its per-capita growth is decreased by an amount proportional to the population of species $j$.
The amount of this decrease in growth is determined by the interaction matrix $A_{ij}$.
In the $S \gg 1$ case, one assumes that the interaction matrix $A_{ij}$ has random entries.
We take $A_{ij}$ to have the statistics:
\begin{equation}
  \begin{aligned}
    \ev{A_{ij}}
    =
    \frac{\mu}{S},
    \qquad
    \cov[A_{ij}, A_{kl}]
    =
    \frac{\sigma^2}{S}
    \qty(
      \delta_{ik} \delta_{jl}
      +
      \gamma\,
      \delta_{il} \delta_{jk}
    )
  \end{aligned}
\end{equation}
This setup means that the variance of each entry is $\sigma^2/S$ and the correlations between the entries are zero except for $\corr[A_{ij},A_{ji}] = \gamma$.
When $\gamma = 1$, the model is symmetric, meaning that $i$ effects $j$ in the same way that $j$ effects $i$.
When $\gamma = -1$, the model is anti-symmetric, meaning that as $i$ inhibits $j$, $j$ promotes $i$.

The random GLV model has a very rich phenomenology including phase transitions, chaotic dynamics, marginal stability, aging, alternative stable states, and more \cite{bunin,cui2024leshoucheslecturescommunity,Roy_2019,dePirey_PRX}.
Properties of the GLV model and its relevance to real ecosystems are an active area of research which has led to the development of novel theoretical tools and fascinating results.

\subsection{The cavity method}

Unlike our simple linear model, we cannot simply solve the equations of motion and integrate out all but one degree of freedom to get the DMFT equation of motion.
Instead, we invoke the cavity method.

\subsubsection{Step 1: Add a species}

The first step is to add a cavity species $i = 0$ to the system:
\begin{equation}
\begin{aligned}
    \dv{N_i}{t}
    &=
    N_i
    \qty(
      1
      -
      N_i
      -
      \sum_{j(\neq i)}
      A_{ij}
      N_j
      +
      A_{i0}
      N_0
      +
      h_i(t)
    ),
    \quad
    i = 1,\ldots,S,
    \\
    \dv{N_0}{t}
    &=
    N_0
    \qty(
      1
      -
      N_0
      -
      \sum_{i=1}^S
      A_{0i}
      N_i
      +
      A_{00}
      N_0
      +
      h_0(t).
    )
\end{aligned}
\end{equation}
In the above, we have added an auxiliary field $h_i(t)$ to the equations of motion for each degree of freedom which we will set to zero at the end.

\subsubsection{Step 2: Treat the new species as a perturbation}

The addition of the new species $N_0$ adds a term $A_{i0} N_0$ to the equations of motion for each species $i = 1,\ldots,S$.
This new term is $O(N^{-1/2})$, so we can treat it as a small perturbation in the $S \gg 1$ limit.
In particular, we can treat it as a shift in the auxiliary field $h_i(t)$:
\begin{equation}
  \begin{aligned}
    h_i(t)
    &\rightsquigarrow
    h_i(t) + A_{i0} N_0(t).
  \end{aligned}
\end{equation}
If we define the dynamical susceptibility matrix:
\begin{equation}
  \begin{aligned}
    \nu_{ij}(t,t')
    \equiv
    \eval{
      \fdv{N_i(t)}{h_j(t')}
    }_{h=0},
  \end{aligned}
\end{equation}
we can relate the dynamics of the system with $S + 1$ species to the dynamics of the system with $S$ species:
\begin{equation}
  \begin{aligned}
    N_i(t)
    &\approx
    N_{i\setminus0}(t)
    +
    \sum_{j=1}^S
    \int_0^t
    \dd{t'}
    \nu_{ij}(t,t')
    A_{j0} N_0(t')
    .
  \end{aligned}
\end{equation}
This linear response approximation becomes exact in the $S \gg 1$ limit.

\subsubsection{Step 3: Obtain the cavity equation of motion}

Substituting this expression into the equation of motion for $N_0(t)$, we get:
\begin{equation}
  \begin{aligned}
    \dv{N_0}{t}
    &=
    N_0
    \qty(
      1
      -
      N_0
      -
      \sum_{i=1}^S
      A_{0i}
      N_{i\setminus0}(t)
      -
      \sum_{i,j=1}^S
      \int_0^t
      \dd{t'}
      \nu_{ij}(t,t')
      A_{0i}
      A_{j0}
      N_0(t')
    )
    .
  \end{aligned}
\end{equation}
Note that we have dropped the $A_{00}$ term because it is $O(1/\sqrt{S})$ and will not contribute in the $S \gg 1$ limit.
This is the cavity equation of motion for the added species $N_0(t)$, and it is exact when we take $S \to \infty$.

\subsubsection{Step 4: Statistics under quenched disorder}

Next, we analyze the statistics of the terms in the cavity equation of motion under quenched disorder and discard $O(N^{-1/2})$ contributions.
The statistics of the direct contribution term are:
\begin{equation}
  \begin{aligned}
    &
    \ev{
      \sum_{i=1}^N
      A_{0i}
      N_{i\setminus0}(t)
    }
    =
    \frac{\mu}{S}
    \sum_{i=1}^S
    N_{i\setminus0}(t)
    =
    \mu m(t),
    \\
    &
    \ev{
      \qty(
      \sum_{i=1}^S
      A_{0i}
      N_{i\setminus0}(t)
      )
      \qty(
      \sum_{j=1}^S
      A_{0j}
      N_{j\setminus0}(t')
      )
    }
    =
    \sum_{i,j=1}^S
    \ev{
      A_{0i}
      A_{0j}
    }
    N_{i\setminus0}(t)
    N_{j\setminus0}(t')
    \\
    &\qquad
    =
    \frac{\sigma^2}{S}
    \sum_{i,j=1}^S
    \delta_{ij}
    N_{i\setminus0}(t)
    N_{j\setminus0}(t')
    =
    \sigma^2
    C(t,t'),
  \end{aligned}
\end{equation}
where we have defined:
\begin{equation}
  \begin{aligned}
    m(t)
    =
    \frac{1}{S}
    \sum_{i=1}^S
    N_{i\setminus0}(t),
    \qquad
    C(t,t')
    =
    \frac{1}{S}
    \sum_{i=1}^S
    N_{i\setminus0}(t)
    N_{i\setminus0}(t')
    ,
  \end{aligned}
\end{equation}
to be the empirical average and auto-covariance of the species abundances.
The back-reaction term has statistics:
\begin{equation}
  \begin{aligned}
    &
    \ev{
      \sum_{i,j=1}^S
      \nu_{ij}(t,t')
      A_{0i}
      A_{j0}
    }
    =
    \gamma \sigma^2 \nu(t,t')
    \\
    &
    \var\qty[
    \sum_{i,j=1}^S
    \nu_{ij}(t,t')
    A_{0i}
    A_{j0}
    ]
    =
    \sum_{i,j,k,l=1}^S
    \nu_{ij}(t,t')
    \nu_{kl}(t,t')
    \cov[
      A_{0i}
      A_{j0}
      ,
      A_{0k}
      A_{l0}
    ]
    \\
    &\qquad
    =
    \frac{\sigma^4}{S^2}
    \sum_{i,j,k,l=1}^S
    \nu_{ij}(t,t')
    \nu_{kl}(t,t')
    \qty(
    \delta_{ik}
    \delta_{jl}
    +
    \gamma^2
    \delta_{il}
    \delta_{jk}
    )
    \\
    &\qquad
    =
    \frac{\sigma^4}{S^2}
    \sum_{i,j=1}^S
    \nu_{ij}(t,t')
    \nu_{ij}(t,t')
    +
    \gamma^2
    \frac{\sigma^4}{S^2}
    \sum_{i,j=1}^S
    \nu_{ij}(t,t')
    \nu_{ji}(t,t')
    \\
    &\qquad
    =
    \frac{\sigma^4}{S}
    \underbrace{
    \Biggl(
    \frac{1}{S}
    % \sum_{i,j=1}^S
    % \nu_{ij}(t,t')
    % \nu_{ij}(t,t')
    \Tr
    \qty[
      \nu(t,t')
      \nu(t,t')^\T
    ]
    +
    \gamma^2
    \frac{1}{S}
    % \sum_{i,j=1}^S
    % \nu_{ij}(t,t')
    % \nu_{ji}(t,t')
    \Tr
    \qty[
      \qty(
        \nu(t,t')
      )^2
    ]
    \Biggr)
    }_{O(1)}
    \\
    &
    \qquad
    =
    O\qty(\frac{1}{S})
    .
  \end{aligned}
\end{equation}
In the above, we have defined:
\begin{equation}
  \begin{aligned}
    \nu(t,t')
    &=
    \frac{1}{S}
    \sum_{i=1}^S
    \nu_{ii}(t,t')
    =
    \frac{1}{S}
    \sum_{i=1}^S
    \eval{
      \fdv{N_i(t)}{h_i(t')}
    }_{h = 0}
    .
  \end{aligned}
\end{equation}
Thus, the back-reaction term is $O(1/\sqrt{S})$ and can be neglected in the $S \gg 1$ limit, meaning we can replace the back-reaction term with its mean.
Putting it together:
\begin{equation}
  \begin{aligned}
    \dv{N_0}{t}
    &=
    N_0
    \qty(
      1
      -
      N_0
      -
      \mu m(t)
      -
      \sigma
      \eta(t)
      -
      \gamma
      \sigma^2
      \int_0^t
      \dd{t'}
      \nu(t,t')
      N_0(t')
    ),
  \end{aligned}
  \label{eq:glv-cavity-eom}
\end{equation}
where $\eta(t)$ is a zero-mean Gaussian process with covariance:
\begin{equation}
  \begin{aligned}
    \ev{\eta(t) \eta(t')}
    &=
    C(t,t')
    .
  \end{aligned}
\end{equation}

\subsubsection{Step 5: Assert self-averaging}

To close the cavity equation of motion, we need to assume that the order parameters $m(t)$, $C(t,t')$, and $\nu(t,t')$ are self-averaging:
\begin{equation}
  \begin{aligned}
    m(t)
    &=
    \frac{1}{S}
    \sum_{i=1}^S
    N_{i\setminus0}(t)
    =
    \ev{N_0(t)}_\mathcal{Q}
    ,
    \\
    C(t,t')
    &=
    \frac{1}{S}
    \sum_{i=1}^S
    N_{i\setminus0}(t)
    N_{i\setminus0}(t')
    =
    \ev{N_0(t) N_0(t')}_\mathcal{Q}
    ,
    \\
    \nu(t,t')
    &=
    \frac{1}{S}
    \sum_{i=1}^S
    \eval{
      \fdv{N_i(t)}{h_i(t')}
    }_{h = 0}
    =
    \ev{\eval{
      \fdv{N_0(t)}{h_0(t')}
    }_{h = 0}
    }_\mathcal{Q}
    .
  \end{aligned}
\end{equation}
To be explicit, we can write an ODE for the dynamical susceptibility in which one writes $\widetilde \nu(t,t') = \fdv{N_0(t)}{h_0(t')}$ and obtains the differential equation:
\begin{equation}
  \begin{aligned}
    \pdv{t}
    \widetilde{\nu}(t,t')
    &=
    \fdv{h_0(t')}
    \dv{N_0(t)}{t}
    \\
    &=
    N_0(t)
    \qty(
      -
      \tilde{\nu}(t,t')
      -
      \gamma
      \sigma^2
      \int_0^t
      \dd{t''}
      \nu(t,t'')
      \widetilde{\nu}(t'',t')
      +
      \delta(t-t')
    )
    \\
    &\qquad
    +
    \widetilde{\nu}(t,t')
    \qty(
      1
      -
      N_0(t)
      -
      \mu m(t)
      -
      \sigma
      \eta(t)
      -
      \gamma
      \sigma^2
      \int_0^t
      \dd{t''}
      \nu(t,t'')
      N_0(t'')
    )
    ,
  \end{aligned}
  \label{eq:glv-nu-self-consistent-ode}
\end{equation}
and then obtains the dynamical susceptibility by averaging over re-samplings of the DMFT stochastic process $N_0(t)$:
\sidenote{
  Note that the notation $\widetilde \nu(t,t')$ represents a single-realization dynamical susceptibility, while $\nu(t,t')$ represents the average over realizations.
  The former is not deterministic and is a stochastic process, while the latter is deterministic.
  \\
  \\
  There is also another way to write a self-consistent equation for $\nu(t,t')$ which takes advantage of Novikov's theorem. See~\cite{Roy_2019} appendix C for details or~\cite{novikov1965functionals} for the original paper.
}
\begin{equation}
  \begin{aligned}
    \nu(t,t')
    =
    \ev{\widetilde{\nu}(t,t')}_\mathcal{Q}
    .
  \end{aligned}
\end{equation}
It is not possible (to the best of my knowledge) to solve these equations analytically, but one can use numerical methods to solve them.
To do this, one makes a guess for the order parameters $m(t)$, $C(t,t')$, and $\nu(t,t')$; solves the stochastic differential equation for $N_0(t)$ many times using the guessed order parameters; takes averages over the sampled trajectories to update the order parameters; and then repeats until convergence.
Ref.~\cite{Roy_2019} provides a detailed description of this procedure and its implementation in Python as well as suggestions to improve the convergence of the numerical method.

\subsubsection{Analyzing the DMFT equation of motion}

We can start ascribing meaning to the terms in the DMFT equation of motion.
We see that the dynamics of a randomly selected species experiences a direct mean-field contribution from the average population of all species, $\mu m(t)$, which decreases the per-capita growth rate of the species.
It additionally experiences fluctuations due to the random interactions with other species, $\sigma \eta(t)$, whose fluctuation size is set by the parameter $\sigma$ and the typical abundance of species. The correlations of $\sigma \eta(t)$ decay the same way as the auto-correlation of the randomly selected species itself.
Finally, the back-reaction term, $\gamma \sigma^2 \int_0^t \dd{t'} \nu(t,t') N_0(t')$, captures how the randomly selected species impacts the entire community and then feeds back into its own dynamics. 
Looking at Eq.~\eqref{eq:glv-nu-self-consistent-ode}, we can see that $\nu(t,t') \geq 0$ which is expected because if we increase the per-capita growth rate artificially (increase $h(t)$), we expect the species abundance to increase.
We additionally see that $\nu(t,t) = \ev{N_0(t)}_\mathcal{Q} = m(t)$ and that $\nu(t,t')$ decreases as $|t - t'|$ increases, so the strength of the self-regulation due to back-reaction is set by $\gamma \sigma^2 m(t)$.
Additionally, when $\gamma > 0$, we see that the back-reaction term indicates there is emergent self-limitation of a species' growth rate due to the interactions with other species.
When $\gamma < 0$, the back-reaction term indicates that the species is promoted by its interactions with other species, which is a form of emergent cooperation.

\subsubsection{Obtaining steady-state solutions}

In the case where the system reaches a steady state, we can treat $N_0(t) = N_0$ as just a constant random variable, $m(t) = m$ a deterministic constant, and $\eta(t) = \eta$ a zero-mean Gaussian random variable with variance $q = C(t,t) = \ev{N_0^2}_\mathcal{Q}$.
For the back-reaction term, we perform the integral to define the integrated response: $\bar{\nu} = \int_0^\infty \dd{t'} \nu(t,t')$.
This gives the static cavity equation:
\begin{equation}
  \begin{aligned}
    0
    =
    N_0
    \qty(
      1
      - 
      N_0
      -
      \mu m
      -
      \sigma
      \sqrt{q}\,
      \xi
      -
      \gamma
      \sigma^2
      \bar{\nu}
      N_0
    ),
  \end{aligned}
\end{equation}
where $\xi$ is a zero-mean unit-variance Gaussian random variable \cite{bunin}.
We note that there are two possible solutions $N_0 > 0$ and $N_0 = 0$.
We will select the $N_0 > 0$ solution whenever it exists because in ecology, we are generally interested in the uninvadable steady-state (meaning the steady-state in which every species that could possibly stably coexist in the system is present).
This means we can write the steady-state solution as:
\begin{equation}
  \begin{aligned}
    N_0
    =
    \max\qty{
      0,
      \frac{
        1 - \mu m - \sigma \sqrt{q}\, \xi
      }{
        1 + \gamma \sigma^2 \bar{\nu}
      }
    }
    .
  \end{aligned}
\end{equation}
Because $\xi$ is a zero-mean Gaussian random variable, this means that the steady-state distribution of species abundances is a truncated Gaussian distribution.

One can proceed to use this expression to write down self-consistent expressions for $m$, $q$, and $\bar{\nu}$, and these self-consistent equations are much easier to handle than the full DMFT equation of motion.
Analyzing the resulting equations additionally reveals the existence of phase transitions in the system, such as a transition from stable steady-state behavior to chaos, alternative-stable states, or unbounded growth \cite{bunin,cui2024leshoucheslecturescommunity,Roy_2019,dePirey_PRX}.
However, we will not reproduce the full analysis here, and we refer the reader to the original literature for more details.

\section{Conclusion}

In this tutorial, we've taken a journey through the core concepts of Dynamical Mean-Field Theory. Our goal was to demystify this powerful framework by applying it to a simple linear model where every step could be justified with exact calculations. We saw how the dynamics of a system with $N$ interacting variables can, in the large $N$ limit, be described by the stochastic dynamics of a single, representative variable. This simplification hinges on the self-averaging property, where macroscopic quantities for a single realization of the system become identical to an average over all possible realizations of the disorder.

We then introduced the cavity method, a more general and powerful technique that does not rely on having an exactly solvable model. By treating a newly added `cavity' particle as a small perturbation, we re-derived the same DMFT equations for our linear model. This illustrated how the cavity method systematically accounts for the crucial `back-reaction' of the system on itself, a feedback loop that is easy to miss with more naive approximations. The key insight is that the cavity particle's dynamics are governed by a stochastic process whose statistical properties, the correlation $C(t,s)$ and response function $\nu(t,s)$, are determined self-consistently by the dynamics of the cavity particle itself.

Finally, we saw how to apply the cavity method to the non-linear generalized Lotka--Volterra model from theoretical ecology. This led us to a DMFT equation that, while not solvable with pen and paper, provides insights into the system's behavior and can be tractably analyzed to explore a rich phase diagram exhibiting stability, chaos, and unbounded growth.

The tools presented here are the foundation for studying a vast array of complex systems in biophysics and beyond, from the chaotic firing of neural networks to the complex energy landscapes of glassy materials. We hope this gentle introduction has provided you with the intuition to explore the world of disordered systems and to apply these methods in your own research.

\textbf{Funding Acknowledgments:} This work was supported by the Fannie and John Hertz Foundation Fellowship and the National Science Foundation Graduate Research Fellowship Program.

\appendix

\newpage
\section{Appendix: Brief introduction to random matrix theory (RMT)\label{sec:brief-intro-rmt}}

Random matrix theory (RMT) is a branch of mathematics and physics that studies the properties of matrices with randomly chosen entries.
It has its origins in nuclear physics, where it was used to model the energy levels of complex atomic nuclei, but has since found applications in many fields including condensed matter physics, number theory, statistics, and machine learning \cite{potters2020first,cui2024leshoucheslecturescommunity}.
In RMT, one often studies the properties of the matrices as they become large, referred to as the thermodynamic limit.
This is because there are many universal properties that emerge in this limit, and if the size of the matrix is small, it might be more appropriate to study the specific matrix rather than its statistical properties.

\subsection{Observables}

When dealing with an $N \times N$ random matrix $A_{ij}$, it might be tempting to ask questions about the distribution of moments like $\ev{\sum_{k=1}^NA_{ik}A_{kj}}$ or $\ev{\sum_{k,l=1}^N A_{ik}A_{kl}A_{lj}}$, but this would result in answers that have size $O(N^2), O(N^3)$, etc. and are not very informative because they can fluctuate a lot from sample to sample.
% Instead, one studies observables that are self-averaging in the thermodynamic limit, meaning that they converge to a deterministic value as $N \to \infty$.
The analogous quantity to the moments of a random variable is the normalized trace:
\begin{equation}
  \begin{aligned}
    \frac{1}{N} \Tr A^k
  \end{aligned}
\end{equation}
For a well-behaved function $f$, one can also compute the normalized trace of $f(A)$ which is the average of $f$ applied to the eigenvalues of $A$:
\begin{equation}
  \begin{aligned}
    \frac{1}{N}
    \Tr f(A)
    =
    \frac{1}{N}
    \sum_{i=1}^N
    f(\lambda_i).
  \end{aligned}
\end{equation}
Both of these quantities are self-averaging in the thermodynamic limit for many common random matrix ensembles, meaning that they converge to a deterministic value as $N \to \infty$.
When $N \to \infty$, the eigenvalues $\lambda_i$ of $A$ are often treated as random samples from a deterministic distribution $\rho(\lambda)$ called the spectral density, which is defined as:
\begin{equation}
  \begin{aligned}
    \rho(\lambda)
    =
    \lim_{N \to \infty}
    \frac{1}{N}
    \sum_{i=1}^N
    \delta(\lambda - \lambda_i),
  \end{aligned}
\end{equation}
where $\delta(x)$ is the Dirac delta function.
This is interpreted as the empirical distribution of the eigenvalues of $A$.
As $N \to \infty$, $\rho(\lambda)$ often converges to a continuous function, and the eigenvalues become dense in a certain interval.
In the case where this is well-defined, we can write:
\begin{equation}
  \begin{aligned}
    \frac{1}{N}
    \Tr f(A)
    &\to
    \int
    \dd{\lambda}
    \rho(\lambda)
    f(\lambda),
  \end{aligned}
\end{equation}
which reveals that the normalized trace is the expectation of $f(\lambda)$ when $\lambda$ is drawn from the spectral density $\rho(\lambda)$.

\subsection{Wigner ensemble and the scaling of entries}

A common random matrix ensemble is the Wigner ensemble, which consists of $N \times N$ symmetric matrices $A$ with entries $A_{ij}$ that are i.i.d. normal random variables with mean zero and variance $\sigma^2/N$ for $i < j$, and $A_{ii} = 0$.
The Wigner ensemble is a prototypical example of a random matrix ensemble that exhibits universal properties in the thermodynamic limit and is analogous to the normal distribution for random variables because there are analogous theorems to the central limit theorem and law of large numbers.

The variance of the entries of the Wigner ensemble is scaled as $\sim 1/N$ because this ensures that the eigenvalues of $A$ remain $O(1)$ as $N \to \infty$.
To see this, consider the variance of the eigenvalue distribution which is given by the normalized trace of $A^2$:
\begin{equation}
  \begin{aligned}
    \var[\lambda]
    \to
    \ev{
    \frac{1}{N}
    \Tr A^2
    }
    =
    \frac{1}{N}
    \sum_{i,j=1}^N
    \ev{
      A_{ij} A_{ji}
    }
    =
    \frac{2}{N}
    \sum_{i,j (i\neq j)}
    \var[A_{ij}]
  \end{aligned}
\end{equation}
Unless $\var[A_{ij}] \sim 1/N$, this quantity would either diverge or vanish as $N \to \infty$.
The fact that with this scaling, quantities like $\frac{1}{N} \Tr f(A)$ converge to $O(1)$ deterministic values as $N \to \infty$ plays an important role in the derivations presented in the main text.

The spectral density of the Wigner ensemble converges to the Wigner semicircle law in the thermodynamic limit:
\begin{marginfigure}
  \includegraphics[width=\marginparwidth]{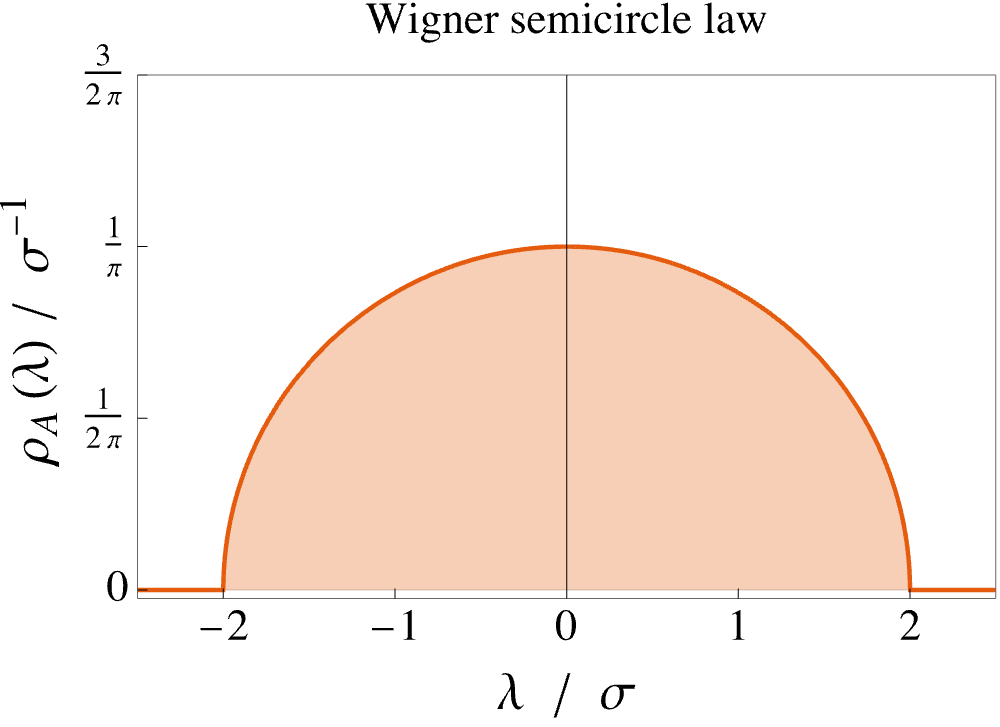}
  \caption{\label{fig:wigner-semicircle}
  The Wigner semicircle law for the spectral density of a symmetric random matrix with i.i.d. normal entries with mean zero and variance $\sigma^2/N$.
  }
\end{marginfigure}
\begin{equation}
  \begin{aligned}
    \rho(\lambda)
    &=
    \begin{cases}
      \frac{1}{2\pi \sigma^2}
      \sqrt{4 \sigma^2 - \lambda^2},
      & \abs{\lambda} < 2\sigma,\\
      0,
      & \abs{\lambda} \geq 2\sigma,
    \end{cases}
  \end{aligned}
\end{equation}
The spectral densities of random matrices often have compact support like this, meaning that there is a maximum and minimum eigenvalue in the thermodynamic limit.
The exceptions to this are ensembles of matrices that become singular in the thermodynamic limit because this corresponds to having very large eigenvalues.
In this appendix, we will derive the Wigner semicircle law, but it is not necessary for understanding the main text.

\subsection{Ginibre ensemble}

The Ginibre ensemble is another common random matrix ensemble which consists of $N \times N$ matrices $Q$ with entries $Q_{ij}$ that are i.i.d. normal random variables with mean zero and variance $\sigma^2/N$.
The Ginibre ensemble is not symmetric, so its eigenvalues are generally complex numbers.
The spectral density of the Ginibre ensemble converges to the circular law in the thermodynamic limit:
\begin{equation}
  \begin{aligned}
    \rho(x + i y)
    &=
    \begin{cases}
      \frac{1}{\pi \sigma^2},
      & \abs{x + i y} < \sigma,\\
      0,
      & \abs{x + i y} \geq \sigma.
     \end{cases}
  \end{aligned}
\end{equation}

\subsection{Ginibre matrices as random rotations}

Recall that a matrix $Q$ which has $Q^\T Q = Q Q^\T = I$ is called orthogonal, and that such matrices represent rotations and reflections in $N$-dimensional space.
If you have a vector $\vb*x \in \mathbb{R}^N$ and you multiply it by an orthogonal matrix $Q$, the resulting vector $Q \vb*x$ has the same length as $\vb*x$.
If you have some matrix $M$ and form $Q M Q^\T$, this corresponds to rotating the basis in which $M$ is expressed.
A Ginibre matrix $Q$ is not orthogonal, but it is approximately orthogonal in the thermodynamic limit in the sense that:
\begin{equation}
  \begin{aligned}
    [Q Q^\T]_{ij}
    &=
    \sum_{k=1}^N
    Q_{ik} Q_{jk}
    \approx
    \frac{\sigma^2}{N}
    \delta_{ij},
  \end{aligned}
\end{equation}
where $\delta_{ij}$ is the Kronecker delta (1 if $i = j$ and 0 otherwise).
This means that multiplying by a Ginibre matrix approximately preserves lengths and angles in high dimensions, so Ginibre matrices can be thought of as random rotations.

This is useful when diagonalizing a random symmetric matrix $M$.
If $M$ is a symmetric matrix, it can be diagonalized as $M = O \Lambda O^\T$ where $O$ is an orthogonal matrix and $\Lambda$ is a diagonal matrix of eigenvalues.
In many cases, the columns of $O$ (which are the eigenvectors of $M$) are called `delocalized' because the entries of each eigenvector are approximately independent normal random variables with variance $1/N$.
It is often a good approximation to treat $O$ as a Ginibre matrix in the thermodynamic limit, which is useful for computing averages over eigenvectors of random matrices.
This means that one can think of many random symmetric matrices as being a random rotation of a diagonal matrix whose entries are sampled from the spectral density.

\subsection{Wick's theorem\label{sec:wick-theorem}}

A very useful tool for computing averages over Gaussian random variables is Wick's theorem, which states that the average of a product of an even number of zero-mean Gaussian random variables can be expressed as a sum over all possible pairings of the variables, where you take the product of the averages of each pair.
For example, if $x_1, x_2, x_3, x_4$ are zero-mean Gaussian random variables, then:
\begin{equation}
  \begin{aligned}
    \ev{x_1 x_2 x_3 x_4}
    &=
    \ev{x_1 x_2} \ev{x_3 x_4}
    +
    \ev{x_1 x_3} \ev{x_2 x_4}
    +
    \ev{x_1 x_4} \ev{x_2 x_3}.
  \end{aligned}
\end{equation}
If there is an odd number of variables, the average is zero.
In the math and statistics literature, this is often called Isserlis' theorem.

\subsection{The Green's function as the moment generating function}

A very useful tool for computing the spectral density of a random matrix is the Green's function, which is defined as:
\begin{equation}
  \begin{aligned}
    G(z)
    &\equiv
    \lim_{N \to \infty}
    \frac{1}{N}
    \Tr
    \qty(z I - A)^{-1}
    =
    \lim_{N \to \infty}
    \frac{1}{N}
    \sum_{i=1}^N
    \frac{1}{z - \lambda_i},
  \end{aligned}
\end{equation}
with $z$ a complex number not in the spectrum of $A$.
Let's see what happens if we expand this in powers of $1/z$:
\begin{equation}
  \begin{aligned}
    G(z)
    =
    \frac{1}{N}
    \sum_{i=1}^N
    \qty(
      \frac{1}{z}
      +
      \frac{\lambda_i}{z^2}
      +
      \frac{\lambda_i^2}{z^3}
      +
      \ldots
    )
  \end{aligned}
\end{equation}
This means that we can compute:
\begin{equation}
  \begin{aligned}
    \frac{1}{N}
    \Tr
    A^k
    &=
    \eval{
      \dv[k+1]{w}
      G(w^{-1})
    }_{w=0}
  \end{aligned}
\end{equation}
Thus, $G(z)$ is the moment generating function for the eigenvalue distribution of $A$.
If we know $G(z)$, we can recover the spectral density using the Stieltjes inversion formula.
To do this, first observe:
\begin{equation}
  \begin{aligned}
    \lim_{\varepsilon \to 0^+}
    \frac{1}{x \pm i \varepsilon}
    &=
    \mp 
    i 
    \pi
    \lim_{\varepsilon \to 0^+}
    \frac{
       \varepsilon
    }{
        \pi(x^2 + \varepsilon^2)
    }
    +
    \lim_{\varepsilon \to 0^+}
    \frac{
      x
    }{
      x^2 + \varepsilon^2
    }
    \\
    &
    =
    \mp i \pi \delta(x)
    +
    \mathrm{P.V.}
    \frac{1}{x},
    \\
    \implies
    &
    \frac{1}{\pi}
    \lim_{\varepsilon \to 0^+}
    \Im
    \frac{1}{x - i \varepsilon}
    =
    \delta(x),
  \end{aligned}
\end{equation}
where P.V. denotes the Cauchy principal value and $\delta(x)$ is the Dirac delta function.
We can use this to extract the spectral density from $G(z)$:
\begin{equation}
  \begin{aligned}
    \rho(\lambda)
    &
    =
    \lim_{N \to \infty}
    \frac{1}{N}
    \sum_{i=1}^N
    \delta(\lambda - \lambda_i)
    =
    \frac{1}{\pi}
    \lim_{N \to \infty}
    \lim_{\varepsilon \to 0^+}
    \Im
    \frac{1}{N}
    \sum_{i=1}^N
    \frac{1}{\lambda - \lambda_i - i \varepsilon}
    \\
    &=
    \frac{1}{\pi}
    \lim_{\varepsilon \to 0^+}
    \Im
    G(\lambda - i \varepsilon)\
    .
  \end{aligned}
\end{equation}
Thus, if we can compute $G(z)$ for complex $z$, we can recover the spectral density $\rho(\lambda)$.

\subsection{Derivation of the Wigner semicircle law}

Here, we'll present one way to derive the Wigner semicircle law using the Green's function.
There are many other ways to get this result that work for more general ensembles of random matrices, but this is one of the simplest.

Consider a Wigner matrix $A$ with entries $A_{ij}$ that are i.i.d. normal random variables with mean zero and variance $\sigma^2/N$ for $i < j$, and $A_{ii} = 0$.
Let $R(z) = (z I - A)^{-1}$ be the resolvent of $A$.
If we start with $(zI - A)R(z) = I$ and take the normalized trace of both sides, we get:
\begin{equation}
  \begin{aligned}
    z G(z)
    =
    1
    +
    \frac{1}{N}
    \Tr
    [A R(z)].
  \end{aligned}
\end{equation}
Let's turn to the second term on the right-hand side.
We can invoke an identity called Stein's lemma: if $Z$ is a zero-mean normal random variable with variance $\var[Z]$, then for any well-behaved function $f$, we have:
\begin{equation}
  \begin{aligned}
    \ev{Z f(Z)}
    =
    \var[Z]
    \ev{f'(Z)}.
  \end{aligned}
\end{equation}
This can be derived by integration by parts.
We can use this to compute:
\begin{equation}
  \begin{aligned}
    \ev{
      \frac{1}{N}
      \sum_{i,j=1}^N
      A_{ij} R_{ji}(z)
    }
    =
    \frac{1}{N}
    \sum_{i,j=1}^N
    \var[A_{ij}] 
    \ev{
      \pdv{R_{ji}(z)}{A_{ij}}
    }
  \end{aligned}
\end{equation}
Returning to $(zI -A)R(z) = I$, differentiating with respect to $A_{ij}$, using the product rule, and solving, we get:
\begin{equation}
  \begin{aligned}
    \pdv{A_{ij}}
    \qty[\frac{1}{zI - A}]_{kl}
    =
    \qty[\frac{1}{zI - A}]_{k i}
    \qty[\frac{1}{zI - A}]_{j l}
  \end{aligned}
\end{equation}
Thus:
\begin{equation}
  \begin{aligned}
    \frac{1}{N}
    \Tr
    [A R(z)]
    &=
    \frac{\sigma^2}{N^2}
    \sum_{i,j=1}^N
    R_{j i}(z)
    R_{j i}(z)
    \\
    &
    =
    \frac{\sigma^2}{N^2}
    \sum_{i,j(i\neq j)}
    R_{j i}(z)
    R_{j i}(z)
    +
    O\qty(
      \frac{1}{N}
    )
    \\
    &
    \approx
    \sigma^2
    \qty(
      \frac{1}{N}
      \Tr R(z)
    )^2
  \end{aligned}
\end{equation}
where we have dropped the $O(1/N)$ terms in the thermodynamic limit.
This gives us the equation:
\sidenote{
% It is no coincidence that this equation is the same as Eq.~\eqref{eq:dmft-eom-cavity-G-resolvant-equation} in the main text (except for a sign difference which is due to the skew-symmetry of the coupling matrix in the main text).
It is no coincidence that $\widetilde{G}(z)$, the Laplace transform of the Green's function for the DMFT equation of motion, follows a very similar equation to $G(z)$ here.
In particular (see Eq.~\eqref{eq:dmft-eom-cavity-G-resolvant-equation} in the main text):
\begin{align*}
  z \widetilde{G}(z)
  =
  1
  -
  \sigma^2
  \widetilde{G}(z)^2
  .
\end{align*}
There is a sign difference because the coupling matrix $A_{ij}$ in the main text is skew-symmetric whereas the Wigner matrix here is symmetric.
There is a very deep connection between the Green's functions of linear systems with random couplings and the spectra of random matrices.
In fact, one can exploit this connection to quickly compute the spectral density of many random matrix ensembles using the cavity method.
To learn more about this, see Ref.~\cite{cuiPRM}.
}
\begin{equation}
  \begin{aligned}
    z G(z) = 1 + \sigma^2 G(z)^2.
  \end{aligned}
\end{equation}
Solving this quadratic equation for $G(z)$, we get:
\begin{equation}
  \begin{aligned}
    G(z)
    =
    \frac{
      z - \sqrt{z^2 - 4 \sigma^2}
    }{
      2 \sigma^2
    },
  \end{aligned}
\end{equation}
where we choose the branch of the square root that ensures $G(z) \sim 1/z$ as $z \to \infty$.
Finally, we can use the Stieltjes inversion formula to recover the spectral density:
\begin{equation}
  \begin{aligned}
    \rho(\lambda)
    &=
    \frac{1}{\pi}
    \lim_{\varepsilon \to 0^+}
    \Im
    G(\lambda - i \varepsilon)
    =
    \begin{cases}
      \frac{1}{2\pi \sigma^2}
      \sqrt{4 \sigma^2 - \lambda^2},
      & \abs{\lambda} < 2\sigma,\\
      0,
      & \abs{\lambda} \geq 2\sigma.
    \end{cases}
  \end{aligned}
\end{equation}
This is the Wigner semicircle law.

\subsection{The skew-symmetric case}

In the main text, we deal with a random coupling matrix $A$ that is skew-symmetric, meaning that $A^\T = -A$.
This means that the eigenvalues of $A$ are purely imaginary.
The matrix $iA$ is Hermitian and has eigenvalues distributed according to the Wigner semicircle law with variance $\sigma^2$.
If $\mu$ is an eigenvalue of $i A$, then $\lambda = -i \mu$ is an eigenvalue of $A$.
Thus, the spectral density of $A$ is just the Wigner semicircle law rotated in the complex plane onto the imaginary axis:
\begin{equation}
  \begin{aligned}
    \rho(x + i y)
    &=
    \delta(x)\,
    \begin{cases}
      \frac{1}{2\pi \sigma^2}
      \sqrt{4 \sigma^2 - y^2},
      & \abs{y} \leq 2\sigma,
      \\
      0,
      & \abs{y} > 2\sigma,
    \end{cases}
  \end{aligned}
\end{equation}
where the Dirac delta $\delta(x)$ indicates that all eigenvalues lie on the imaginary axis.

\section{Appendix: Brief review of stationary Gaussian processes\label{sec:brief-review-stationary-gaussian-processes}}

A Gaussian process is a collection of random variables which are indexed by a continuous parameter (usually time or space) such that any finite subset of these random variables has a multivariate normal distribution.
Gaussian processes are useful for modeling functions and time series because they can represent a wide range of behaviors while still being mathematically tractable.
For example, Gaussian processes can be used to model biological processes like neural activity or population dynamics.
In this appendix, we'll provide a brief review of Gaussian processes and some special properties of those whose statistical properties are invariant under time translation, called stationary Gaussian processes.

\subsection{Definition}

Let's first consider some random variables $X_1, \ldots, X_n$ that have a multivariate normal distribution with mean vector $\mu_i$ and covariance matrix $\Sigma_{ij}$.
This means that their joint probability density function is given by:
\begin{equation}
  \begin{aligned}
    p(X_1, \ldots, X_n)
    \propto
    \exp
    \qty(
      -\frac{1}{2}
      \sum_{i,j=1}^n
      (X_i - \mu_i)
      \Sigma_{ij}^{-1}
      (X_j - \mu_j)
    ),
  \end{aligned}
  \label{eq:multivariate-normal-pdf}
\end{equation}
where $\Sigma_{ij}^{-1}$ is the inverse of the covariance matrix.
Now, let's consider $i = 1,2,\ldots, n$ indexing time points $t_1, t_2, \ldots, t_n$ which are linearly spaced with spacing $\Delta t$.
This means we could interpret $X_i$ as the value of some random function $X(t)$ at time $t_i$.
We might then ask what happens as we take the limit $\Delta t \to 0$ and $n \to \infty$ while keeping $n \Delta t = T$ fixed.
In this limit, we can think of $X(t)$ as a random function defined for all $t \in [0, T]$.
Eq.~\eqref{eq:multivariate-normal-pdf} then becomes:
\begin{equation}
  \begin{aligned}
    p(\{X(t)\})
    \propto
    \exp
    \qty(
      -\frac{1}{2}
      % \sum_{i,j=1}^n
      \int_0^T
      \dd{t}
      \int_0^T
      \dd{s}
      (X(t) - \mu(t))
      C^{-1}(t,s)
      (X(s) - \mu(s))
    )
  \end{aligned}
\end{equation}
where $C(t,s)$ is the covariance function defined as:
\begin{equation}
  \begin{aligned}
    C(t,s)
    &= \ev{(X(t) - \mu(t))(X(s) - \mu(s))},
  \end{aligned}
\end{equation}
and $C^{-1}(t,s)$ is the unique function such that:
\begin{equation}
  \begin{aligned}
    \int_0^T
    \dd{t}
    \int_0^T
    \dd{s}
    C^{-1}(t,s) C(s,t')
    &= \delta(t - t'),
  \end{aligned}
\end{equation}
where $\delta(t - t')$ is the Dirac delta function.
When $p(X_1,\ldots, X_n)$, becomes $p(\{X(t)\})$ we call it a probability density functional (meaning that it takes a function as an input and outputs a probability density), and we say that $X(t)$ is a Gaussian process.

\subsection{Stationary Gaussian processes}

A stationary Gaussian process is a Gaussian process whose mean and covariance are invariant under time translation.
This means that the mean function $\mu(t)$ is constant and the covariance function $C(t, s)$ depends only on the time difference $t - s$:
\begin{equation}
  \begin{aligned}
    \mu(t) &= \mu,\qquad
    C(t, s) &= C(t - s).
  \end{aligned}
\end{equation}
These sorts of processes appear in systems that are in equilibrium or steady state, where the statistical properties do not change over time.
In the case where $\lim_{t \to \infty} C(t) = 0$, the process is not correlated with itself at long times.
When this is the case, averages over long time intervals can be used to estimate ensemble averages.

\subsection{Power spectral density and Wiener--Khinchin theorem}

A useful tool for analyzing stationary Gaussian processes is the power spectral density (PSD), which describes how the variance of the process is distributed over different frequencies.
The PSD of a function $X(t)$ is defined as the average squared magnitude of its Fourier transform:
\begin{equation}
  \begin{aligned}
    S_X(\omega)
    =
    \ev{
      \lim_{T \to \infty}
      \frac{1}{T}
      \abs{
        \int_0^\infty
        \dd{t}
        X(t)
        e^{-i \omega t}
      }^2
    },
  \end{aligned}
\end{equation}
where the expectation is over realizations of the process.
The Wiener--Khinchin theorem states that the PSD is the Fourier transform of the auto-covariance function:
\begin{equation}
  \begin{aligned}
    S_X(\omega)
    &=\lim_{T\to\infty}
    \frac{1}{T}\abs{\int_{0}^{T}\dd{t}X(t)e^{-i\omega t}}^2\\
    &=\lim_{T\to\infty}
    \frac{1}{T}\int_{0}^{T}\int_{0}^{T}\dd{t}\dd{s}
    C(t-s)e^{-i\omega(t-s)}\\
    &=
    \int_{-\,\infty}^{\infty}
    C(\tau)\,e^{-i\omega\tau}\,\mathrm{d}\tau
    .
    \end{aligned}
\end{equation}

% \section{Two additional calculations}

% Here, we present two additional calculations that are mentioned in the main text but not essential for understanding the cavity method or the DMFT equation of motion.

% % \subsection{The generating functional of $\Phi_\alpha(t)$ is self-averaging\label{sec:generating-functional-self-averaging}}

% \subsection{There are no anomalous fluctuations in $\ev{\Phi_\alpha(t)}_\text{sh}$\label{sec:no-anomalous-fluctuations}}

\nocite{*}
% \begin{widetext}
{\small \bibliography{references}}
% \end{widetext}

\end{document}